\documentclass[fleqn,usenatbib,useAMS]{mnras}

\usepackage{graphicx}	
\usepackage{amsmath}	
\usepackage{amssymb}	
\usepackage{multicol}        
\usepackage{bm}		
\usepackage{epstopdf}
\pdfoutput=1

\newcommand{\be}{\begin{equation}}
\newcommand{\ee}{\end{equation}}
\newcommand{\bea}{\begin{eqnarray}}
\newcommand{\eea}{\end{eqnarray}}

\newcommand{\bdm}{\begin{displaymath}}
\newcommand{\edm}{\end{displaymath}}

\newcommand\ab {{\overline a} }
\newcommand\de{{^\circ} }

\usepackage[T1]{fontenc}
\usepackage{ae,aecompl}

\title[Secular models for planets in coorbital 3D motion]{Secular models and Kozai resonance for planets in coorbital non-coplanar motion}
\author[Giuppone \& Leiva]{\large{C. A. Giuppone$^{1,2}$\thanks{E-mail: cristian@oac.unc.edu.ar}, A.M.Leiva$^{1}$} \\
$^{1}$Universidad Nacional de C\'ordoba, Observatorio Astron\'omico, Laprida 854, X5000BGR C\'ordoba, Argentina \\
$^{2}$Universidad Nacional de C\'ordoba, Observatorio Astron\'omico, IATE, Laprida 854, X5000BGR C\'ordoba, Argentina 
}

\date{MNRAS accepted. Last updated 2016 April 27; in original form 2015 December 11}

\pubyear{2015}

\begin{document}
\label{firstpage}
\pagerange{\pageref{firstpage}--\pageref{lastpage}}
\maketitle

\begin{abstract}

In this work, we construct and test an analytical and a semianalytical secular models for two planets locked in a coorbital non-coplanar motion, comparing some results with the case of restricted three body problem. 

The analytical average model replicates the numerical N-body integrations, even for moderate eccentricities ($\lesssim$ 0.3) and inclinations ($\lesssim10^\circ$), except for the regions corresponding to quasi-satellite and Lidov-Kozai configurations. Furthermore, this model is also useful in the restricted three body problem, assuming very low mass ratio between the planets. We also describe a four-degree-of-freedom semianalytical model valid for any type of coorbital configuration in a wide range of eccentricities and inclinations. 

{Using a N-body integrator, we have found that the phase space of the General Three Body Problem is different to the restricted case for inclined systems, and establish the location of the Lidov-Kozai equilibrium configurations depending on mass ratio. We study the stability of periodic orbits in the inclined systems, and find that apart from the robust configurations $L_4$, $AL_4$, and $QS$ is possible to harbour two Earth-like planets in orbits previously identified as unstable $U$ and also in Euler $L_3$ configurations, with bounded chaos.}
\end{abstract}


\section{Introduction} \label{intro}

The three body problem has been studied since decades, particularly with more interest in the coorbital problem. The coorbital problem or 1:1 mean motion resonance (1:1 MMR) occurs when considering a central star and two planets. The period of the planets is almost the same, although the resonance acts avoiding collisions between the bodies. During the last years, several approaches were developed to find new types of regular orbits for this resonance and, in particular, surface of sections in parametric spaces \citep{Hadjidemetriou_etal_2009,Hadjidemetriou_etal_2011},
semi-analytical models \citep{Giuppone_etal_2010}, and analytical models \citep{Robutel_2013} were used. 

Efforts have been made to determine the possibility of the detection of coorbital planets through the radial velocity signal \citep{Giuppone_etal_2012, Dobrovolskis_2013, Leleu_2015}, transit detection \citep{Ford_2006} or transit timing variations in the case that one or both planets transit the stellar disc  \citep{Vokrouhlicky_2014, Haghighipour_etal_2013, Ford_2007}. Although we still do not know details of dominant formation and evolutionary processes of these planetary systems, as well as their type, a general discussion has been established about whether or not the planets can be captured in the MMR 1:1. 

Particularly, in non-coplanar case, we think that it is important to compare the general problem to the restricted problem because these results can be applied on our own Solar System. For example, the  dynamical structure of the coorbital region provides a possible origin for coorbital satellites of the planets. As pointed by \citet{Namouni_1999} and \citet{Mikkola_etal_2006} transitions from Horseshoe or Tadpole orbits to quasi satellite orbits can be thought as a transport mechanism of distant coorbiting objects to a state of temporary or permanent capture around the planet. Once trapped, additional mechanisms provide subsequent permanent capture, for example, collisions with other satellites, mass growth of the planet and the drag of the circunplanetary nebula. This model can be useful even in the formation of the Janus-Epimetheus system through collisions. Recently, \citet{Morais_2016} showed that resonant capture in coorbital motion is present for both prograde and retrograde orbits. 

\begin{figure}
 \centering
 \mbox{\includegraphics[width=8.0cm]{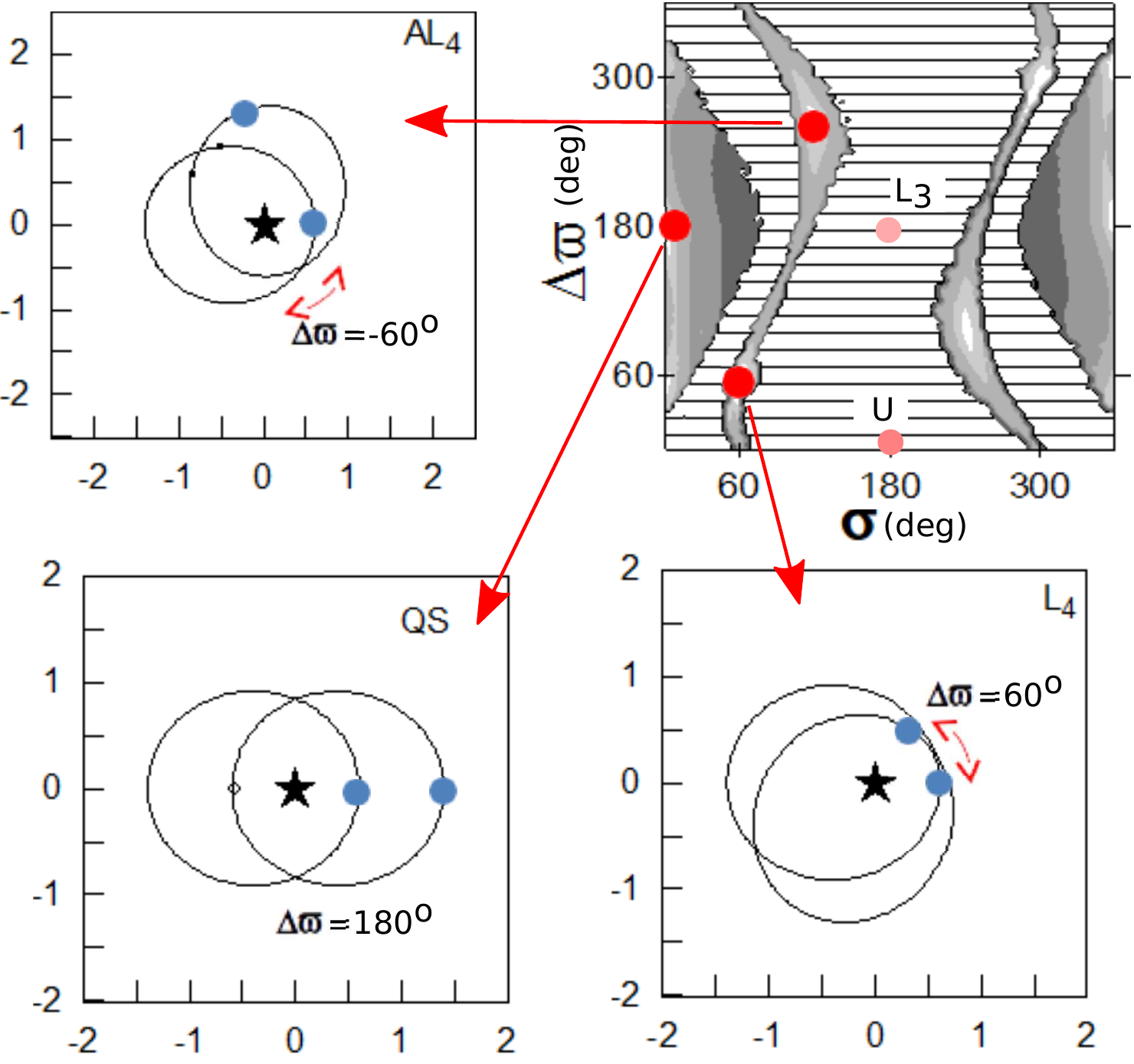}}\\
\caption{Top right-hand panel shows a dynamical map on the plane ($\sigma,\Delta\varpi$) with the colour scale representing the oscillation amplitude of $\sigma$. Initial osculating elements correspond to two Jupiter planets orbiting a 1 $M_\odot$ star at 1 au with initial osculating eccentricities $e_i=0.4$. The gray scale indicates the amplitude of oscillation for $\sigma$ and the dashed region corresponds to unstable configurations. In the remaining panels we identify the three periodic orbits $QS$, $L_4$ and $AL_4$ and plot their representation in the plane ($x,y$) with the star at the origin. Initial conditions for both planets are shown with blue circles, with $m_1$ located along the $x$-axis. Both axis directions are fixed.}
 \label{Fig.schema}
\end{figure}

Classical celestial mechanics books \citep{brouwer_clemence_1961,Moulton_1914} deal with Lagrangian equilibrium points and the orbits around them in the context of the Restricted Three Body Problem (RTBP): Horseshoe (HS) and Tadpole (TP) orbits. However, some other equilibrium orbits were identified, recently. As far as we know, three different kind of periodic orbits can be found in the averaged general three body problem. It is convenient to describe the configurations with two angles $(\sigma,\Delta\varpi) = (\lambda_2-\lambda_1,\varpi_2-\varpi_1)$, where $\lambda_i$ are mean longitudes and $\varpi_i$ are longitudes of pericentre of the planets. Apart from the well known equilateral configurations, located at the classical equilibrium Lagrangian points ($L_4$ and $L_5$) with angles $(\sigma,\Delta\varpi) = (\pm 60^\circ,\pm 60^\circ)$, Quasi-Satellite ($QS$) orbits and Anti-Lagrangian orbits ($AL_4$ and $AL_5$) are present. For low eccentricities, Anti-Lagrangian orbits are located at $(\sigma,\Delta\varpi) = (\pm 60^\circ,\mp 120^\circ)$. One anti-Lagrangian solution $AL_i$ is connected to the corresponding $L_i$ solution through the $\sigma$-family of periodic orbits in the averaged system (the solutions with zero-amplitudes of the $\sigma$--oscillation). The $QS$ orbits are characterized by oscillations around a fixed point, which is always located at $(\sigma,\Delta\varpi) = (0^\circ,180^{\circ})$, independently on the planetary mass ratio and eccentricities. In the top right-hand panel of Figure \ref{Fig.schema}, we construct a dynamical map with a grey scale indicating the amplitude of oscillation of $\sigma$ on the plane $(\sigma,\Delta\varpi)$ identifying the equilibrium orbits. Each of the other plots shows the orbital representation of some configurations in ($x, y$) astrocentric Cartesian coordinates. We focus our attention on $L_4$ and $AL_4$ configurations, because $L_5$ and $AL_5$ configurations are dynamically equivalent to the formers \citep[see][]{Giuppone_etal_2010,Hadjidemetriou_etal_2009}. {Additionally, in the Figure \ref{Fig.schema} we marked with light circles the location of Euler configuration, $L_3$, and the center of unstable family $U$ studied by \citet{Hadjidemetriou_etal_2009} and afterwards related with the $L_3$ configuration by \citet{Robutel_2013}. We pay special attention to both configurations at the final section. Note that $L_3$ is located at $(\sigma,\Delta\varpi) = ( 180^\circ, 180^\circ)$, while unstable configuration $U$ is located at $(\sigma,\Delta\varpi) = (180^\circ, 0^\circ)$.}

In Section \ref{model} we present the Hamiltonian analytical model with elliptic expansions and explore the validity of the average model. Also, we compare the results with direct N-body integrations. In Section \ref{semi} we introduce the average semianalytical model for the three-body problem in non-coplanar case, extending previous results, and compare to numerically filtered integrations. Following, in Section \ref{3d}, we focus on the study of 3D equilibrium orbits, particularly on the Lidov-Kozai resonance with the different models. Finally, conclusions are presented in Section \ref{Conclusions}.


\section{Analytical model} \label{model}

Classical expansions of the disturbing function do not converge when the semi-major axis ratio is $\simeq 1$, and consequently they are not appropriate to model the coorbital resonance. Then, our intention is to give an easy handle Hamiltonian to describe the motion within this resonance. 
We consider a system of two planets with masses $m_{i}$ moving around a star with mass $m_{0}$ with inclinations lower than 90$^\circ$. We not include additional planets neither dissipative forces. Each planetary orbit is described by six orbital elements: semi-major axis $a$, eccentricity $e$, inclination $i$, longitude of pericentre $\varpi$, mean longitude in orbit $\lambda$, and longitude of the node $\Omega$. Alternatively we can use the argument of pericentre $\omega=\varpi-\Omega$, mean anomaly $M=\lambda-\varpi$, and true anomaly $f$.
 
We write the Hamiltonian following \citet{laskar_robutel_1995}, using a canonical set of variables introduced by Poincar\'e with astrocentric positions of the planets $\mathbf{r}_{i}$, and barycentric momentum vectors $\mathbf{p}_{i}$. The pairs $(\mathbf{r}_{i},\mathbf{p}_{i})$ form a canonical set of variables with the Hamiltonian given by

\begin{equation}
 H = H_{0} + U + T
\label{hamil}
\end{equation}

\noindent Here, $H_{0}$ is the Keplerian part (sum of the independent Keplerian Hamiltonians), $U$ is the direct part, and $T$ is the kinetic part of the Hamiltonian, written in terms of the canonical variables $(\mathbf{r}_{i}, \mathbf{p}_{i})$ as

\begin{equation}
\begin{split}
 H_{0}         =& -\sum_{i=1}^{N}\left(\frac{p_{i}^{2}}{2\beta_{i}} - \frac{m_{0}m_{i}}{||\mathbf{r_{i}}||}\right),\\
 {U}  =& -\mathcal{G}\sum_{i,j=1 \; i \ne j}^{N}\frac{m_{i}m_{j}}{\Delta_{ij}},\\
 {T}  =& \sum_{i,j=1 \; i \ne j}^{N}\frac{\mathbf{p}_{i}\cdot\mathbf{p}_{j}}{m_{0}},
\label{hamiltonian}
\end{split}
\end{equation}

\noindent where $\mathcal{G}$ is the gravitational constant, \ $\beta_{i} = m_{0}m_{i}/(m_{0} + m_{i})$ and $\Delta_{ij} = ||\mathbf{r}_{i} - \mathbf{r}_{j}||$. 

In the three-body problem, the barycentric momenta, $\mathbf{p}_{i}$, are related to the heliocentric velocities, $\dot{\mathbf{r}}_{i}$,  by the following expressions
\begin{eqnarray}
\mathbf{p}_1&=& \frac{m_1}{m_0+m_1+m_2} \left[ (m_0+m_2) \dot{\bf{r}}_1-m_2 \dot{\bf{r}}_2 \right], \\\nonumber
\mathbf{p}_2&=& \frac{m_2}{m_0+m_1+m_2} \left[ (m_0+m_1) \dot{\bf{r}}_2-m_1 \dot{\bf{r}}_1 \right]. 
\end{eqnarray}
For the planetary case $m_i \ll m_0$, then
\begin{eqnarray}
\mathbf{p}_1&\simeq& m_1 \dot{\bf{r}}_1 +  \frac{m_1 m_2 }{m_0} \left( \dot{\bf{r}}_1 - \dot{\bf{r}}_2\right)\\\nonumber
\mathbf{p}_2&\simeq& m_2 \dot{\bf{r}}_2 +  \frac{m_1 m_2 }{m_0} \left( \dot{\bf{r}}_2 - \dot{\bf{r}}_1\right). 
\end{eqnarray}

\noindent The distance, $\Delta$, between the planets is
\begin{equation}
 \Delta^2=r_1^2+r_2^2-2 r_1 r_2 \cos \phi
\end{equation}
\noindent being $\phi$, the angle between the vectors $\mathbf{r}_1$ and $\mathbf{r}_2$, 
\begin{equation}
\begin{split}
   \cos \phi &= \frac{\mathbf{r}_1 \cdot \mathbf{r}_2}{r_1 r_2}\\
\end{split}
\end{equation}
\begin{figure*}
\centering
 \mbox{\includegraphics[width=0.68\columnwidth]{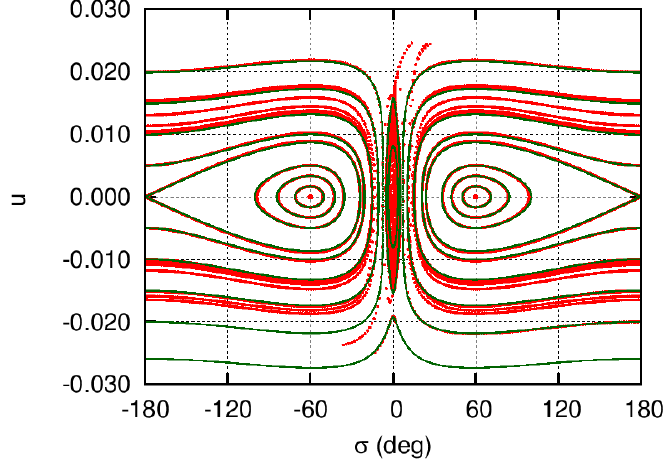}} \mbox{\includegraphics[width=0.68\columnwidth]{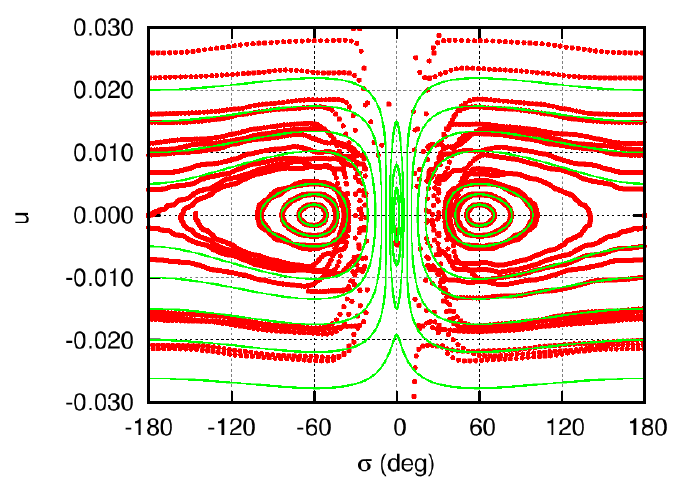}} \mbox{\includegraphics[width=0.68\columnwidth]{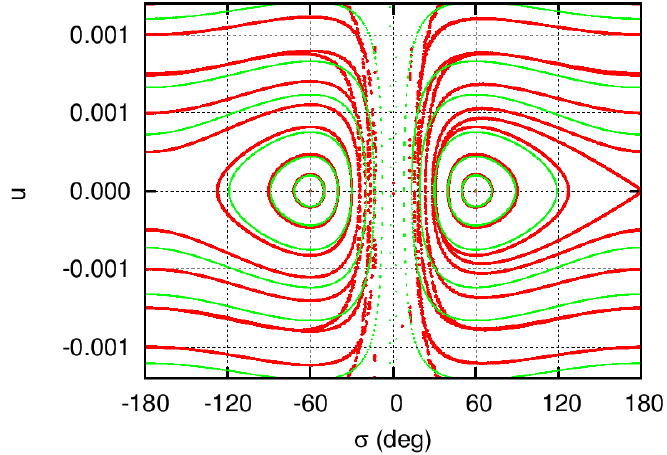}} \\
 \mbox{\includegraphics[width=0.68\columnwidth]{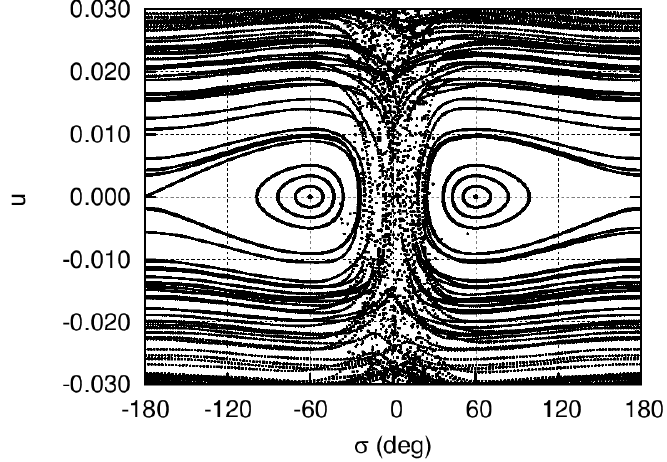}}  \mbox{\includegraphics[width=0.68\columnwidth]{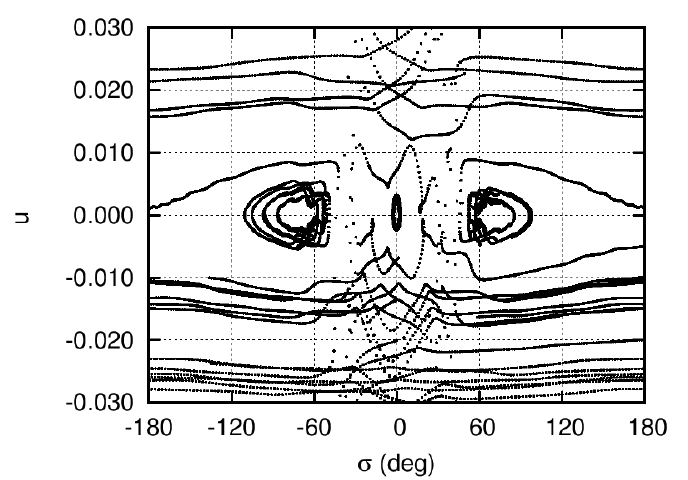}}  \mbox{\includegraphics[width=0.68\columnwidth]{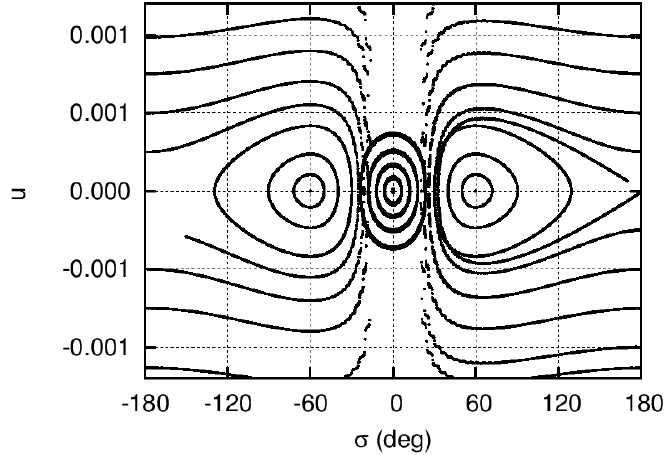}}  \\
\caption{Top row. Phase space described by initial conditions integrated using $\mathcal{H}$ and $\mathcal{H}_{00}$ in the plane ($u,\sigma$). Jupiter-like planets in quasi circular orbits (left panel), Jupiter-like planets in eccentric orbits (middle panel) and Earth-like planets in eccentric orbits (right panel). Bottom-row. The same initial conditions integrated with the N-body code for 400 years. See text for details.}
 \label{fig1}
\end{figure*}

We then expand positions and velocities in eccentricities, $e_j$, and inclinations, $s_j=\sin (i_j/2)$, obtaining an analytic expansion for the Hamiltonian $\mathcal{H}$. Also, we keep the coefficients up to the order $\mathcal{O}(e_j^2), \mathcal{O}(s_j^2)$, and $\mathcal{O}(m_j)$. Then, we integrate over the fast angle $\lambda_1+\lambda_2$, recovering the averaged analytical Hamiltonian $\mathcal{H}_2$ as
\begin{equation}
\begin{split}
\mathcal{H}_2   &= \mathcal{H}_{00} + \mathcal{G} m_1 m_2 \mathcal{H}_{22}, \\
\mathcal{H}_{00} &=-\frac {\beta_1 \mu_1}{2 a_1}  - \frac{\beta_2 \mu_2}{2 a_2} + 
\mathcal{G} m_1 m_2 \left( \frac{\cos\sigma}{\sqrt{a_1 a_2}} - \frac{1}{\tilde{\Delta}}\right), \\
\mathcal{H}_{22} &= H_{2000} \ (e_1^2 + e_2^2) + H_{1100} \ e_1 e_2  \\
   &+H_{0020} \ (s_1^2 + s_2^2)  +H_{0011} \ s_1 s_2,  \\
\end{split}
\end{equation}
\noindent where 
\begin{equation}
\begin{split}
      \mu_i   &= \mathcal{G} (m_0 + m_i), \\
      \sigma &= \lambda_2-\lambda_1,\\
       \tilde{\Delta}  &= \sqrt{a_1^2+a_2^2-2 a_1 a_2 \cos (\sigma),}
\end{split}
\end{equation}
$\mathcal{H}_{00}$ has zero-order terms in eccentricities and inclinations, and $\mathcal{H}_{22}$ has order two terms, formally:
\begin{equation}
\begin{split}
H_{2000} = & - \frac{\cos (\sigma)}{2 \sqrt{a_1 a_2}}  \\
           & + \frac{a_1 a_2}{8 \tilde{\Delta}^5} \left[ 4 \cos (\sigma) (a_1^2+a_2^2) + a_1 a_2 (5 \cos (2 \sigma)-13) \right], \\
H_{1100}  = & \frac{\cos ({\Delta}\varpi-2 \sigma)}{\sqrt{a_1a_2}} + \frac{\gamma}{\tilde{\Delta}^5}, \\
\gamma    = & -a_1 a_2( a_1^2 + a_2^2) \cos ({\Delta}\varpi-2 \sigma) -\frac{a_1^2 a_2^2}{8}  \\
            & [\cos ({\Delta}\varpi-3 \sigma)- 26 \cos ({\Delta}\varpi-\sigma)+ 9 \cos ({\Delta}\varpi+\sigma)], \\
H_{0020} &= \left(\frac{a_1 a_2}{\tilde{\Delta}^3}-\frac{1}{\sqrt{a_1 a_2}}\right) \cos (\sigma) \\
H_{0011} &= 2 \left( \frac{1}{\sqrt{a_1 a_2}}-\frac{a_1 a_2 }{\tilde{\Delta}^3} \right)  \cos (\Omega_2-\Omega_1-\sigma).
\label{eq.expansion}
\end{split}
\end{equation}
This expression for $\mathcal{H}_2$ is equivalent to the one reported in \citet{Robutel_2013}, but avoiding the Complex {notation}. We also want to remark that, due to the D'Alembert rules only even powers of eccentricities and inclinations are present in $\mathcal{H}_2$.  

The first-order average Hamiltonian, given by the expression $\mathcal{H}_{2}$, is not valid in the region of $QS$ because the fast angle ($\lambda_1+\lambda_2$) has a similar period to that of the resonant one ($\sigma$) \footnote{Recently, \citet{Robutel_2015} have proposed a valid rigorous average method in this region.}. 

The integrable approximation $\mathcal{H}_{00}$, associated to the circular and planar resonant problem, was used by some authors to study the motion inside the resonance because it should provide qualitative information about the {system dynamics}. However, this approximation is inadequate to describe the real dynamics of the planets, even in some simple cases. We put in evidence this fact comparing the integrations projected in the plane ($u,\sigma$), {using the analytic expansion $\mathcal{H}$ with the results from the integrable approximation $\mathcal{H}_{00}$; being} 
\begin{equation}
       u   =\frac{\sqrt{\mu_1 \mu_2}} {\sqrt{\mathcal{G} m_0}} \frac{\beta_1 \beta_2}{(\beta_1 + \beta_2)} 
\frac{(\sqrt{a_1}-\sqrt{a_2})} {(\beta_1 \sqrt{\mu_1 a_1}+\beta_2 \sqrt{\mu_2 a_2})}
\end{equation}
the dimensionless non canonical action-like variable. 

In Figure \ref{fig1} we compare the evolution of initial conditions in the plane ($\sigma,u$) using the integrable approximation $\mathcal{H}_{00}$, the analytical expansion $\mathcal{H}$, and N-body simulations. From left to right initial conditions corresponds to two Jupiter-like planets in coplanar quasi circular orbits ($e_i=0.01$), in eccentric orbits ($e_i=0.15$), and two Earth-like planets in eccentric orbits ($e_i=0.15$). Initial conditions are set for $\sigma=2^\circ, 60^\circ, 180^\circ$, \text{and} $300^\circ$ for different $u$ values around zero. Consequently, the semi-major axes are
\be 
 a_i = \ab\left( 1  + (-1)^{i+1} \frac{\beta_1+\beta_2}{\beta_j}\sqrt{\frac{\mu_0}{\mu_j}  } u\right)^2
 \label{eq:aJu}
 \ee 
were the parameter $\ab$ is the mean value around which the semi-major axes oscillate, $\ab=1$ (see \citealp{Robutel_2013}). Also, the initial conditions for $\Delta\varpi$ are set according to the nearest value of the equilibrium solutions, namely $\sigma \simeq 0^\circ \rightarrow \Delta\varpi=180^\circ$, and $\sigma \simeq \pm 60^\circ \rightarrow \Delta\varpi = \pm 60^\circ$. The top row shows integrations given by the analytical $\mathcal{H}$ (red dots) and $\mathcal{H}_{00}$ (green lines), while the bottom row shows the same initial conditions, but integrated with a full N-body code. {Strictly speaking, our figures depict a projection of the orbital elements on the phase-space portraits. In order to draw a formal parallel between numerically computed phase-space portraits and their analytic counterparts, a numerical averaging process must be appropriately carried out over the rapidly varying angles. However, for the purposes of this work, we shall loosely refer to these plots as phase-space portraits, since their information content is almost identical.} 

At bottom Frame of Figure \ref{fig1} we can see that for Jupiter-like planets only small-amplitude Tadpole orbits are stable. The remaining conditions are highly unstable (can be seen as sparsely points) and neither $QS$ nor $HS$ exist for more than a few orbits. Then, we set a threshold to stop the integrations when the mutual distance between bodies is smaller than the sum of their mutual radius (assuming Earth or Jupiter radius, depending on the case) or if they exhibit a chaotic behaviour changing their configuration. We also find transitions from $HS$ or Tadpole orbits to $QS$ orbits. Moreover, for quasi circular orbits ($e_i=0.01$), it is evident that the integrable approximation (top row) $\mathcal{H}_{00}$ is not good to describe the real dynamics, and that the inclusion of lower-order terms of eccentricities present in $\mathcal{H}$ are enough to destabilise the system. Furthermore, only Tadpoles orbits around $L_4$ and $L_5$ remain stable (top left and middle panels of Fig \ref{fig1}). 
At right-hand panel of Fig \ref{fig1}, with moderate initial eccentricities but planetary masses very small, $m_i/m_0 = 3 \times 10^{-6}$, the dynamics predicted by the integrable approximation $\mathcal{H}_{00}$ are similar to $\mathcal{H}$; however, the $QS$ region is only present in the N-body integrations (bottom right panel). 

\begin{figure}
\centering
 \mbox{\includegraphics[width=\columnwidth]{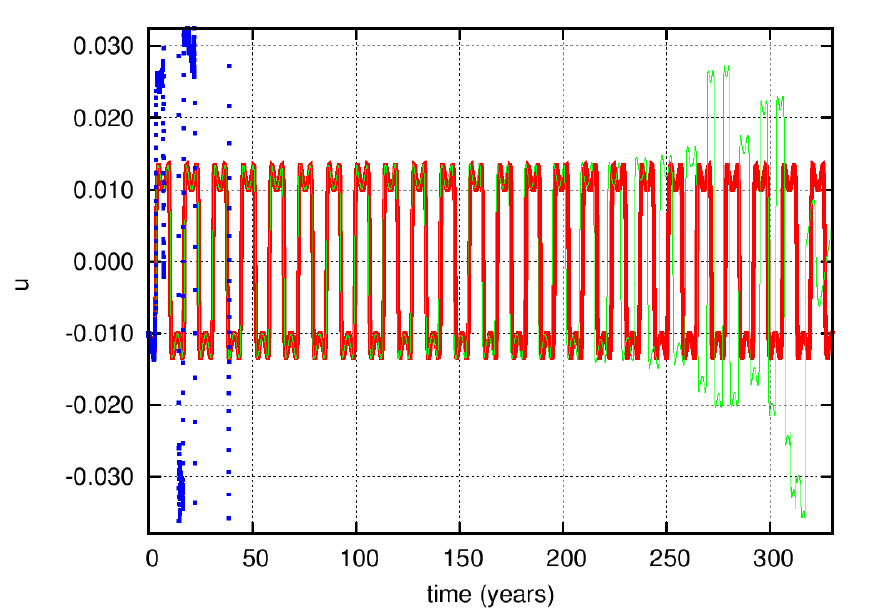}}  \\
\caption{Evolution of $u$ for a coplanar Horseshoe pair of Jupiter planets with initial condition at ($u,\sigma$)=($-0.01,180^\circ$), using $\mathcal{H}_{00}$ (thick red line), $\mathcal{H}$ (green line) and N-body integrations (blue dots). The inclusion of lower-order terms of the eccentricities rapidly excites the system causing the disruption of the resonance (time $\simeq$ 250 periods). The N-body simulation rapidly evidences the chaotic nature of this configuration (time $\simeq$ 5 periods).}
 \label{fig-exp}
\end{figure}

{To understand what happen in $HS$ configuration, the Figure \ref{fig-exp} shows an example of variation of $u$ with time, integrated with different models, for a Jupiter pair of planets. The inclusion of eccentric terms is responsible for instability, even if the initial conditions belong to quasi circular orbits ($e_i=0.01$).}

\begin{figure}
\centering
 \mbox{\includegraphics[width=\columnwidth]{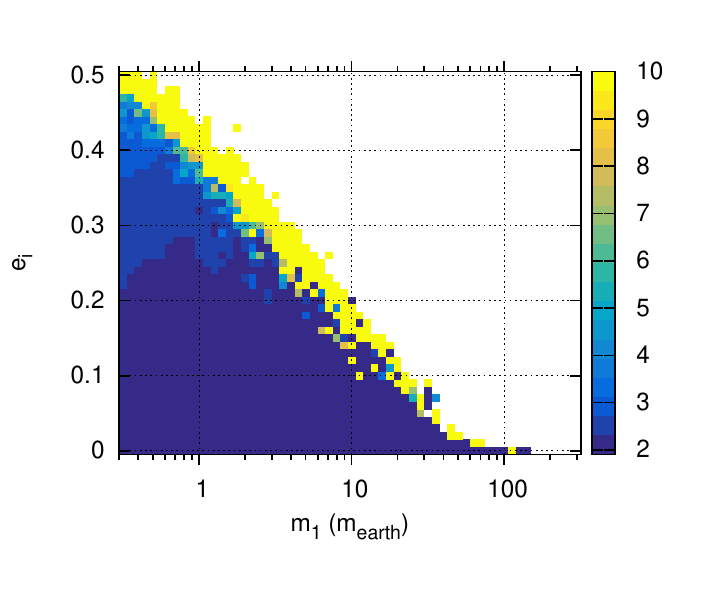}}\\   \vspace{-3em}
 \mbox{\includegraphics[width=\columnwidth]{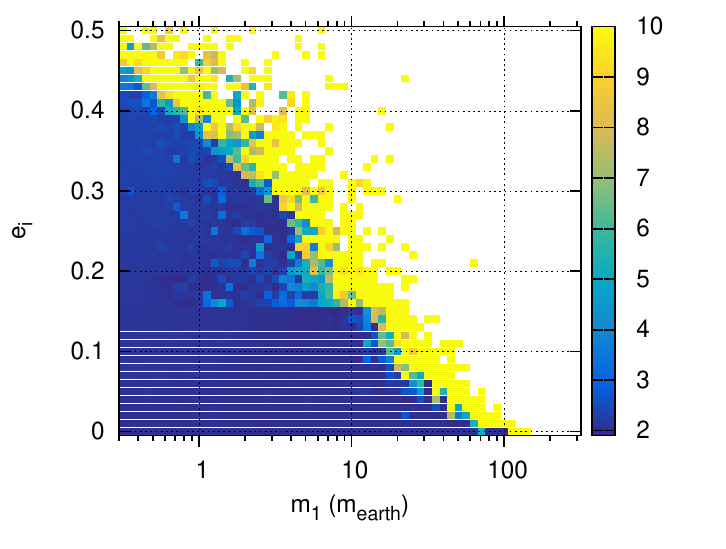}}\\
\caption{{Stability of $HS$ orbits in the plane of osculating initial conditions ($m_1, e_i$) with ($\sigma,\Delta\varpi$)=($60^\circ,60^\circ$). Semi-major axis initial values are taken from Eq.\ref{eq:aJu}, setting $u = 1.2 U_3$. The colour code indicates the value of ${<}Y{>}$. Strongly chaotic systems or systems that quit the coorbital resonance before the integration stops are marked with white dots. All coloured orbits survive for at least $10^5$ periods. Long term integrations show that slow chaotic orbits (${<}Y{>} \gtrsim 5$) survive from $5 \times 10^5$ to $8 \times 10^6$ periods, while unstable conditions (in white) not survive for more than $2 \times 10^3$ periods. Initial conditions correspond to coplanar configurations (top panel), and initial mutual inclinations $J=15^\circ$ (bottom panel).}}
 \label{fig-limit}
\end{figure}
 \begin{figure}
 \centering
\mbox{\includegraphics[width=7.0cm]{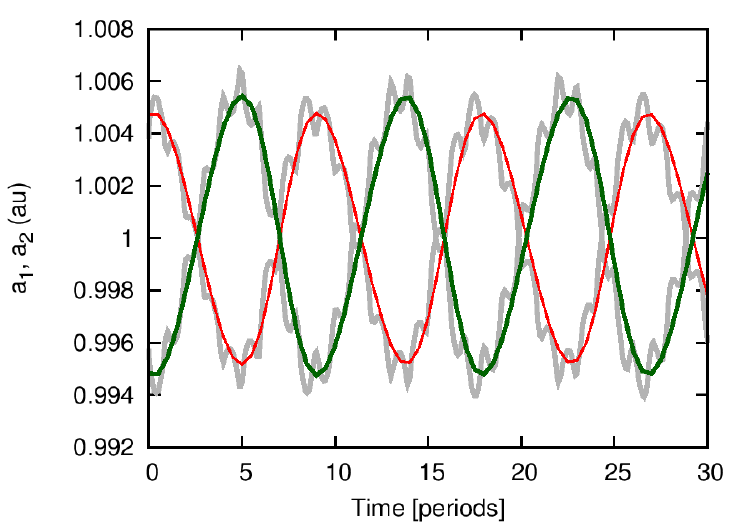}} \\
\mbox{\includegraphics[width=7.0cm]{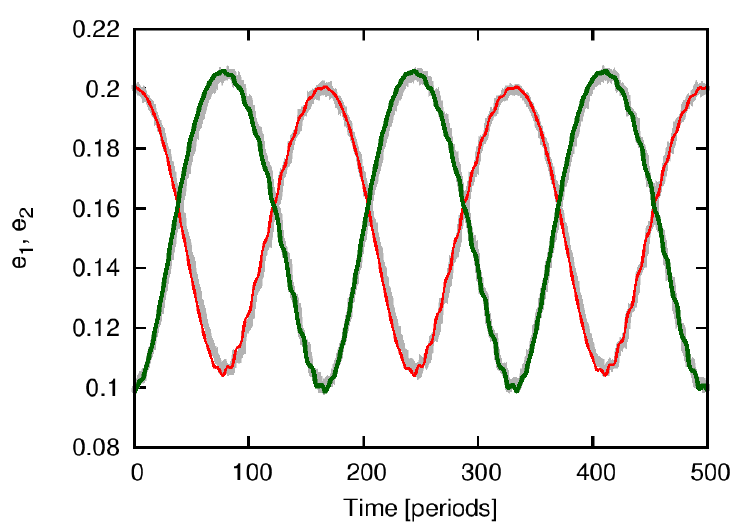}} \\
\mbox{\includegraphics[width=7.0cm]{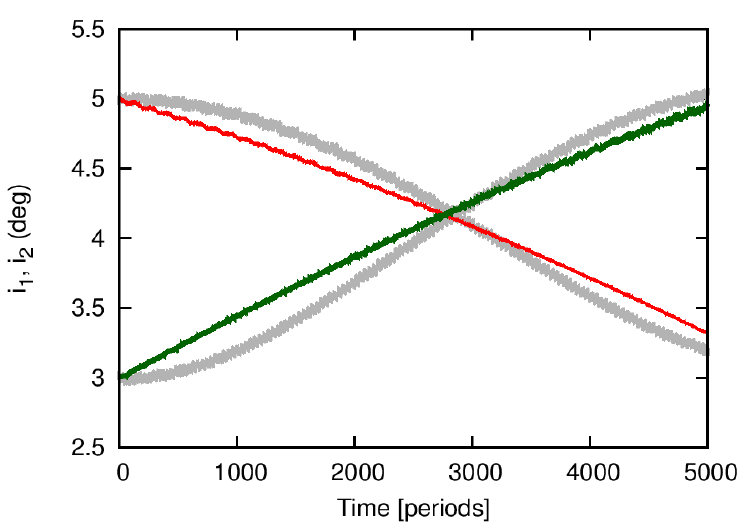}} \\
 \caption{Variation of orbital elements with time using the analytical $\mathcal{H}_2$ model compared with a N-body integration. Amplitudes coincide perfectly and the frequencies were adjusted by hand (see text). Initial conditions from Table \ref{tab1} for the $L_4$ case.}
 \label{figL4}
\end{figure}

\begin{figure}
 \centering
\mbox{\includegraphics[width=7.0cm]{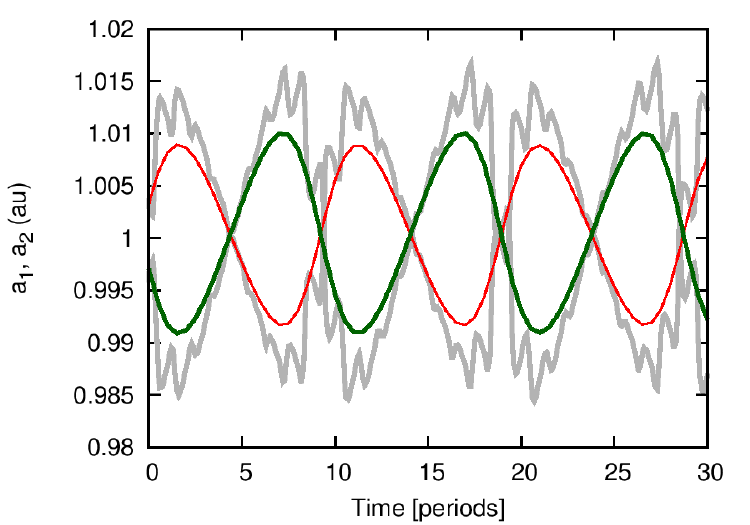}}\\
\mbox{\includegraphics[width=7.0cm]{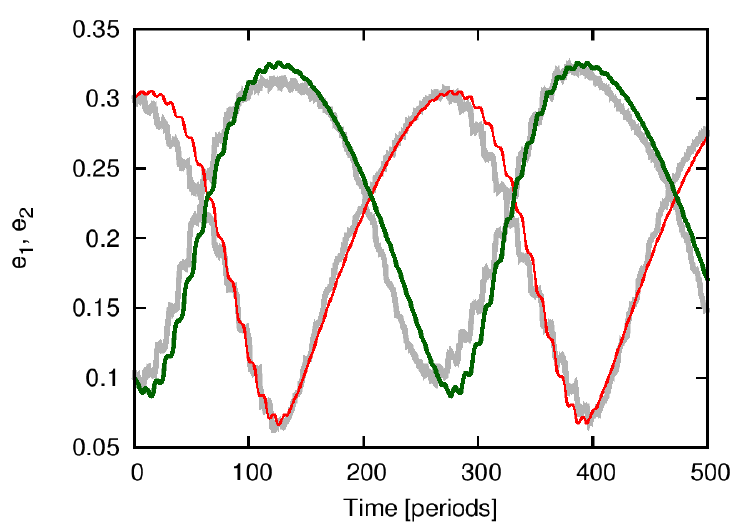}}\\
\mbox{\includegraphics[width=7.0cm]{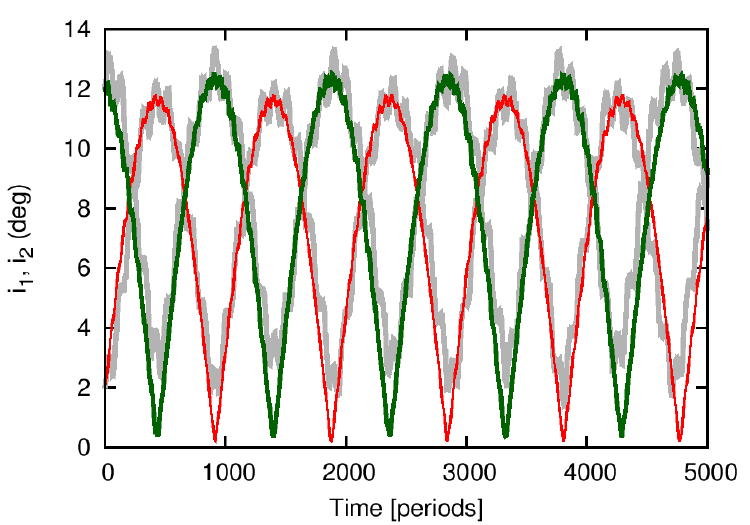}}\\
\caption{Variation of orbital elements with time using the analytical $\mathcal{H}_2$ model compared with a N-body integration. Initial conditions from Table \ref{tab1} correspond to the $AL_4$ case.}
 \label{figAL4}
 \vspace{-2em}
\end{figure}

\begin{figure}
 \centering
\mbox{\includegraphics[width=7.0cm]{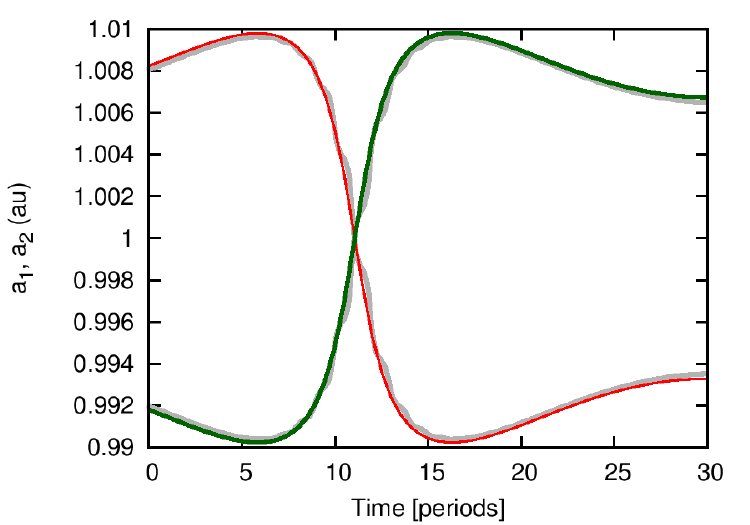}}\\
\mbox{\includegraphics[width=7.0cm]{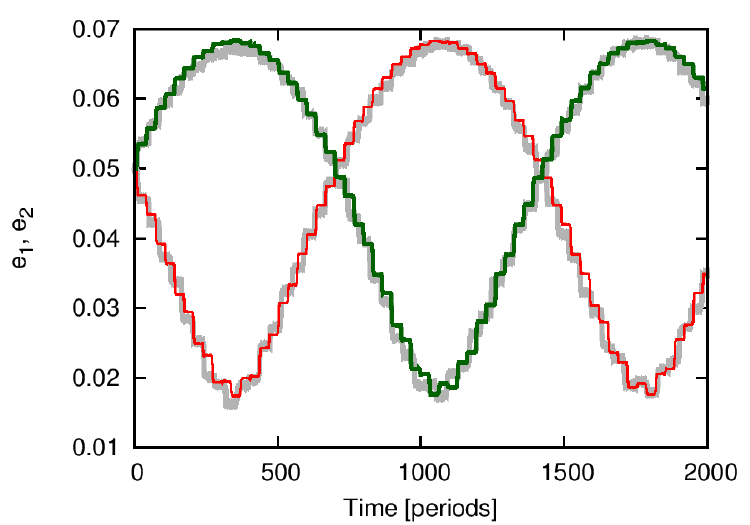}}\\
\mbox{\includegraphics[width=7.0cm]{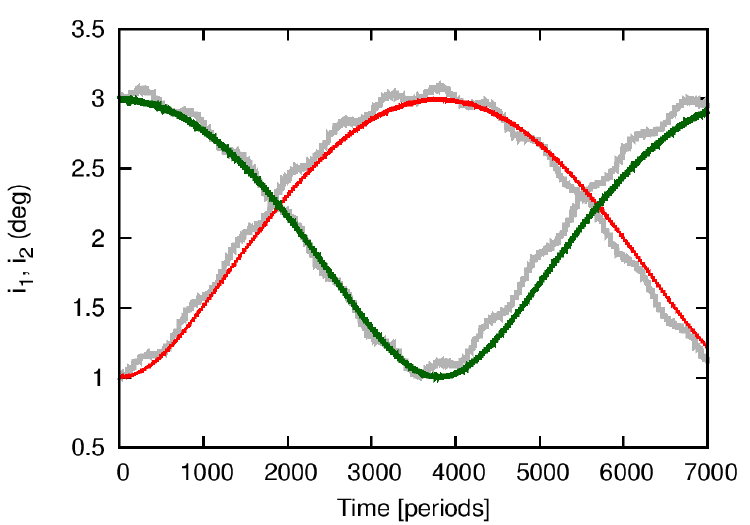}}\\
 \caption{Variation of orbital elements with time using the analytical $\mathcal{H}_2$ model compared with a N-body integration. Initial conditions from Table \ref{tab1} correspond to the $HS$ case.}
 \label{figHS}
 \vspace{-1.5em}
\end{figure}

To study the $HS$ configuration we use the results from \citet{Robutel_2013}, that estimated the size of $HS$ ($U_1$) and $TP$ ($U_3$) region as:

\begin{align}
  U_1 & = \frac{3^{1/6}}{2^{1/3}}\frac{m_1 m_2}{m_0^{1/3}(m_1+m_2)^{5/3}}\nonumber \\
  U_3 & = \frac{2^{1/2}}{3^{1/2}}\frac{m_1 m_2}{m_0^{1/2}(m_1+m_2)^{3/2}}\nonumber\\
  \frac{U_1}{U_3} &=\frac{3^{2/3}}{2^{5/6}} \left(\frac{m_0}{m_1+m_2}\right)^{1/6}.
  \label{eq.sizereg}
\end{align}
Thus, the ratio $\tfrac{U_1}{U_3}$ give us the size of the $HS$ region relative to the $TP$ region. As masses decrease, the relative size increases, but the absolute size is more reduced. 

{\citet{Laughlin_Chambers_2002} mentioned that the $HS$ configuration is not stable for planets more massive than $0.4 {M_J}$ ($\sim$ 100 $M_{\oplus}$) for quasi circular orbits ($e_i=0.01$). Recently,  \citet{Leleu_2015} have shown that the $HS$ configuration is stable for systems with masses lower than $\sim$ 30 $M_{\oplus}$ ($e_i=0.05$). Thus, Setting the initial conditions very close to the value of $U_3$ (1.2 $U_3$) we numerically integrate the three-body problem for different masses ($m_1=m_2$) and initial eccentricities ($e_1=e_2$), and calculate their Mean Exponential Growth factor of Nearby Orbits ${<}Y{>}$ to analyse their chaoticity \citep[i.e. MEGNO,][]{cincotta_2000}. Figure \ref{fig-limit} shows the values of ${<}Y{>}$ for $5 \times 10^4$ periods for coplanar orbits ($J=0^\circ$) and initially mutual inclined orbits ($J=15^\circ$)\footnote{When both planets have masses, it is convenient to work with mutual inclination $J$, defined as $\cos J$=$\cos i_1 \cos i_2 + \sin i_1 \sin i_2 \cos(\Omega_1-\Omega_2)$  (deduced from spherical trigonometry, \citealp[see][pg. 408]{Moulton_1914}).}. In the figure, we can identify the allowed maximum mass values in function of their initial eccentricities for $HS$ planets. These values agree with other authors results regarding to coplanar orbits. We run long-term numerical simulations (10 Myr) for selected initial conditions (specially for $e_i>0.3$) and the initial conditions with ${<}Y{>} \gtrsim 5$ did not survive, maybe due to the long-term diffusion that destabilize the coorbital systems on a time scale that varies from $5 \times 10^5$ to $8 \times 10^6$ periods \citep[see][]{Paez_2015}. Generally, the inclined systems ($J=15^\circ$) can survive for more periods; however, they are strongly chaotic, and those orbits with $e_i\gtrsim0.15$ are frequently transition orbits ($HS-QS$).} 


We have tested the second-order averaged Hamiltonian, $\mathcal{H}_2$, setting the initial conditions near equilibrium configurations with moderate eccentricities ($e_j<0.3$) and mutual inclinations ($J<12^\circ$). Moreover, the mean initial Poincar\'e orbital elements were calculated using a low pass FIR digital filter \citep{Carpino_etal_1987} to eliminate all periodic variations with a period smaller than $3$ years. We have selected initial conditions from Table \ref{tab1} to illustrate the orbital evolution, and the results are shown in Figures \ref{figL4}, \ref{figAL4}, and \ref{figHS}. We can see a  perfect agreement between the N-body integration and the $\mathcal{H}_2$ model for $L_4$, $AL_4$, and the $HS$ configurations respectively. We resolve the Hamiltonian equations using 5 and 6 degrees of freedom, i.e. equations (\ref{eq.poincare}) and (\ref{eq.coorbital}), and the results are the same. 

\begin{figure}
 \centering
\mbox{\includegraphics[width=7.0cm]{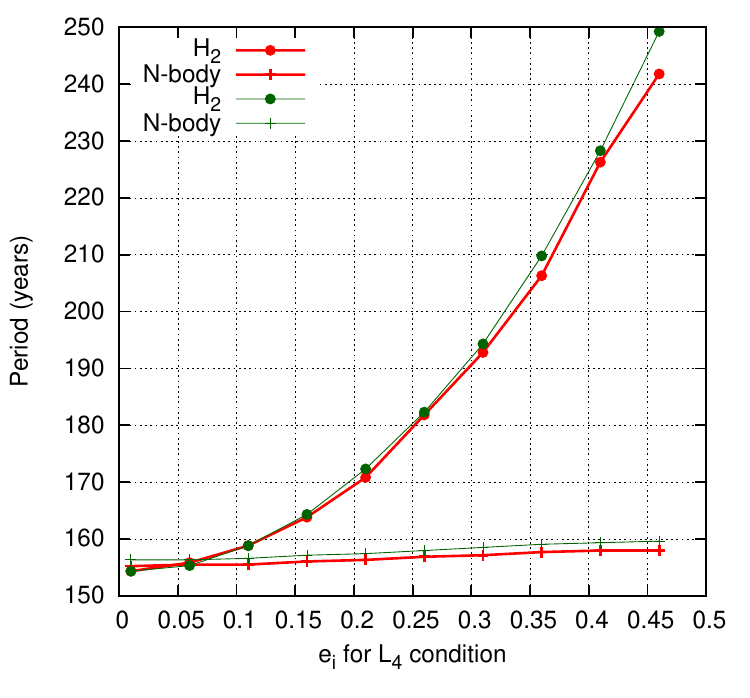}}\\
\mbox{\includegraphics[width=7.0cm]{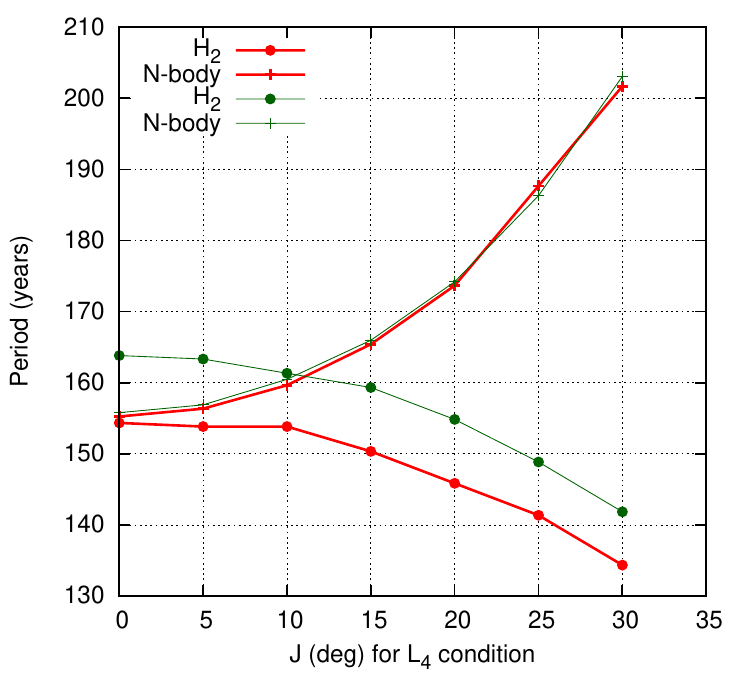}}\\
 \caption{Secular period calculated using the $\mathcal{H}_2$ model (circles) compared with the N-body integration (crosses). Initial osculating angles correspond to the $L_4$ configuration. Top panel. Thick lines have coplanar initial conditions, while thin lines have initial value $J=15^\circ$. Bottom panel. Thick lines have quasi circular initial conditions ($e_i=0.01$), while thin lines have initial values $e_i=0.15$.}
 \label{figper}
 \vspace{-3em}
\end{figure}

We must remark that the integrations with the $\mathcal{H}_2$ model modify the period of the orbital elements. {As a consequence, the secular frequencies sometimes depend on the initial values of $e$ and $i$. Thus, except for very small $e$ and $i$, the secular frequencies are poorly approximated, which is a problem for the study of the resonances (inside the coorbital resonance), and especially the Lidov-Kozai resonance}. For the initial conditions chosen for Figures \ref{figL4}, \ref{figAL4} \& \ref{figHS}, the periods of the eccentricities are 20\% longer than those determined with the N-body integrations. When we modify the initial inclinations, the periods can be even four times the real ones. {To show this, in top panel of Figure \ref{figper} we show the secular periods calculated with $\mathcal{H}_2$ and N-body filtered integrations varying the initial eccentricities and two different initial mutual inclinations ($J=0^\circ$  and $J=15^\circ$), while in bottom panel we set the initial eccentricities at $e_1=e_2=0.01$ and $e_1=e_2=0.15$ for different mutual inclinations. The secular frequencies almost do not depend on the initial values of $e$ in the N-body integrations. For near circular orbits and planar orbits the secular frequencies are well approximated by $\mathcal{H}_2$. When the eccentricity increases the frequencies are poorly determined by the $\mathcal{H}_2$ model. On the contrary, when we fixed the initial eccentricities at $e=0.01$ for different mutual inclinations, neither the N-body simulations nor the $\mathcal{H}_2$ model have constant secular frequencies (bottom panel of Figure \ref{figper}).  }

\begin{table}
\begin{center}
\begin{tabular}{c|c|c|c|c|c|c|c}
 & $\sigma$ & $\Delta\varpi$ & $e_1$ & $e_2$ & $i_1$ & $i_2$  \\
\hline
 $L_4$  & 60 &   60 & 0.2 & 0.1 & 5 & 3 \\
 $AL_4$   & 60 &   240  & 0.3 & 0.1 & 2 & 12 \\
  HS  & 240 &  240 & 0.05 & 0.05 & 1 & 3 \\
   $QS$      &  0   &   180  &  0.45 & 0.45 & 1 & 3    \\ 
\end{tabular} 
\caption{Osculating Poincar\'e initial conditions near the stable periodic solutions in the ($\sigma, \Delta \varpi$) plane. All conditions have all angles in degrees, $m_1=1 M_{J}$, $m_2=0.9 M_{J}$ , $a_1$=1.0038 au, and $a_2$=0.995784 au. $HS$ has $u=0.002$ and masses $m_i=12.5 M_\oplus$. \label{tab1} }
\end{center}
\end{table} 


\section{Semianalytical model}\label{semi}
In order to extend the study of the system to the whole parameter space (e.g. planetary masses, eccentricities, inclinations, etc.), it is useful to construct a semi-analytical model for the coorbital motion. We followed the ideas for the 3D models in other resonances (e.g. \citealp{beauge_michtchenko_2003}) extending the study of coplanar coorbital model developed in \citet{Giuppone_etal_2010}.

Our model involves two main steps: first, a transformation to adequate resonant variables; second, a numerical averaging of the Hamiltonian with respect to short-period terms. Both procedures are detailed below.

We begin introducing the usual mass-weighted Poincar\'e canonical variables \citep[e.g][]{Laskar_1990} for each planet with mass $m_i$:
\bea
\begin{array}{rcll}
\lambda_1 & ; & L_1 = \beta_1 \sqrt{\mu_1 a_1} & × \\ 
\lambda_2 & ; & L_2 = \beta_2 \sqrt{\mu_2 a_2} & × \\ 
p_1=-\varpi_1  & ; & P_1 = L_1-G_1 = L_1\left(1-\sqrt{1-e_1^2}\right) & ×\\
p_2=-\varpi_2  & ; & P_2 = L_2-G_2 = L_2\left(1-\sqrt{1-e_2^2}\right) & ×\\
q_1=-\Omega_1  & ; & Q_1 = G_1-H_1 & ×\\
q_2=-\Omega_2  & ; & Q_2 = G_2-H_2 & ×\\
\label{eq.poincare}
\end{array}
\eea
where $\mu_i = \mathcal{G}(m_0+m_i)$, $G_i=L_i \sqrt{1-e_i^2}$, and $H_i=G_i \cos(i_i)$.

For the initial conditions in the vicinity of coorbital motion, we define the following set of resonant canonical variables $(R_1,R_2,S_1,S_2,T_1,T_2,\sigma,\Delta \varpi,s_1,s_2,t_1,t_2)$, where the new angles and actions are

\begin{align}
   \sigma & =\lambda_2-\lambda_1 & & R_1=\tfrac{1}{2}(L_2-L_1)  & × \nonumber \\
  \Delta\varpi &=p_1-p_2       & & R_2 =\tfrac{1}{2}(P_1-P_2) & × \nonumber\\
  s_1 &= \lambda_1+\lambda_2+p_1+p_2 & & S_1 =\tfrac{1}{2}(L_1+L_2) & × \label{eq.coorbital}\\
  s_2 &= -(p_1+p_2)+(q_1+q_2)        & & S_2 =\tfrac{1}{2}(L_1+L_2-P_1-P_2) & × \nonumber\\
  t_1 &= q_1-q_2       & & T_1 =\tfrac{1}{2}(Q_1-Q_2)                 & × \nonumber\\ 
  t_2 &= -(q_1+q_2)    & & T_2 =\tfrac{1}{2}(H_1+H_2) & × \nonumber
\end{align}
given that
\begin{align}
  a_1 &=  \frac{(S_1-R_1)^2}{\mu_1{\beta_1}^2} \label{eq.a1a2} &
  a_2 &=  \frac{(S_1+R_1)^2}{\mu_2{\beta_2}^2} × \nonumber\\
\end{align}

As we know, a generic argument, $\varphi$, of the disturbing function can be written as:
\be
\label{pphi}
\varphi = j_1\lambda_1+j_2\lambda_2+j_3\varpi_1+j_4\varpi_2+j_5\Omega_1+j_6\Omega_2,
\ee
where $j_k$ are integers. In terms of the new angles, the same argument may be written as:
\begin{equation}
\label{Dal}
\begin{split} 
{2}\varphi = & (j_2-j_1)\sigma + (j_4-j_3)\Delta\varpi + (j_1+j_2) s_1 + \\
          & (\sum_{k=1}^4 j_k) s_2 + (j_6-j_5) t_1 + (\sum_{k=1}^6 j_k) t_2.
\end{split}
\end{equation}

\begin{figure}
 \centering
\mbox{\includegraphics[width=\columnwidth]{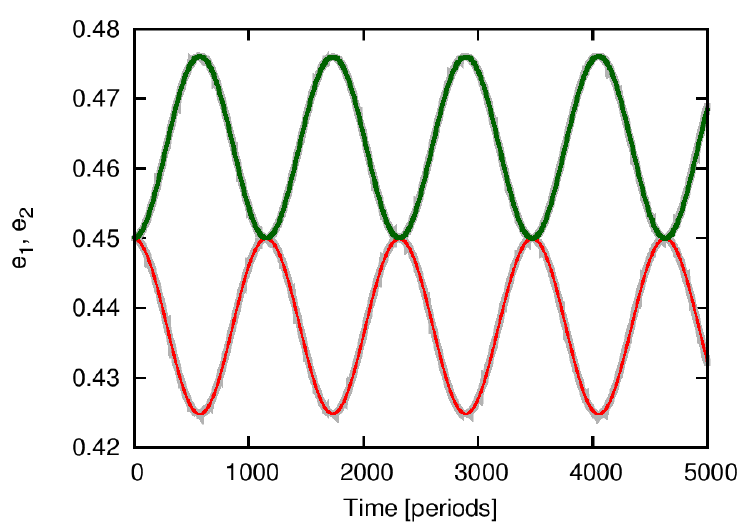}} 
\mbox{\includegraphics[width=\columnwidth]{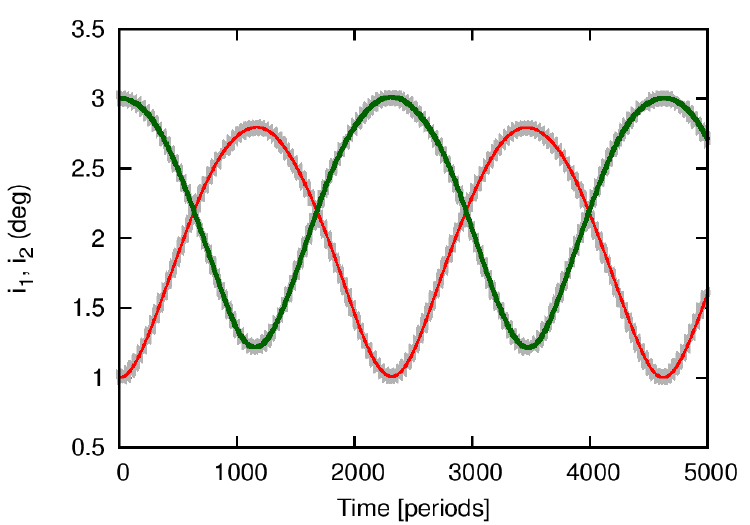}} 
 \caption{Time variation of eccentricities and inclinations using the semianalytical model, $\bar{\mathcal{H}}$, compared with a filtered N-body integration for the $QS$ condition from Table \ref{tab1}.}
 \label{fig.inteQS}
\vspace{-1.5em}
\end{figure}

\begin{figure*}
 \centering
\mbox{\includegraphics[width=0.51\columnwidth]{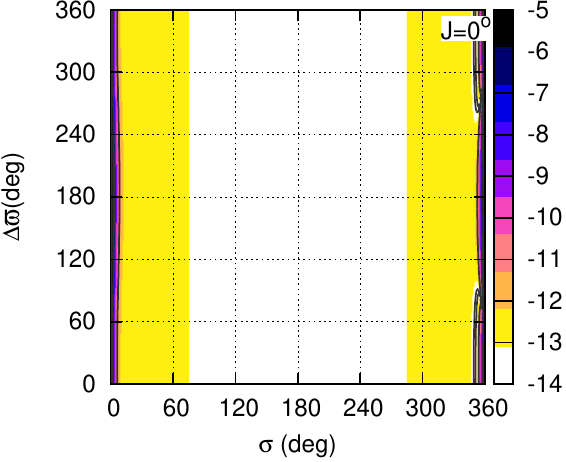}} \mbox{\includegraphics[width=0.51\columnwidth]{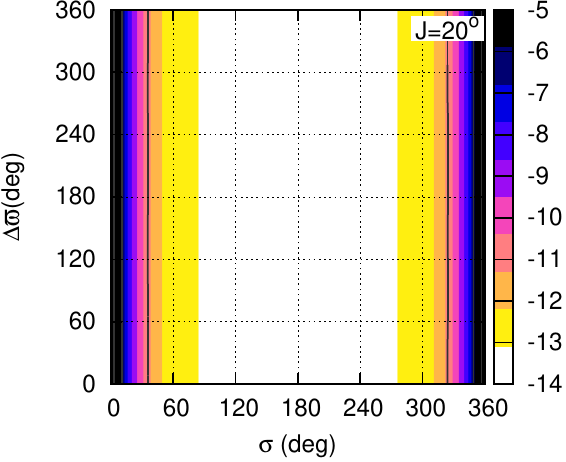}} \mbox{\includegraphics[width=0.51\columnwidth]{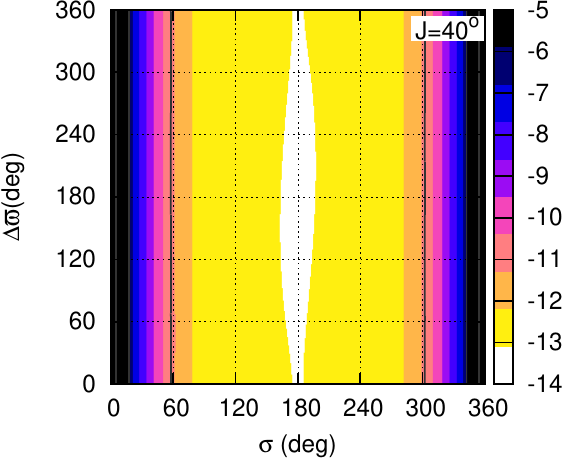}} \mbox{\includegraphics[width=0.51\columnwidth]{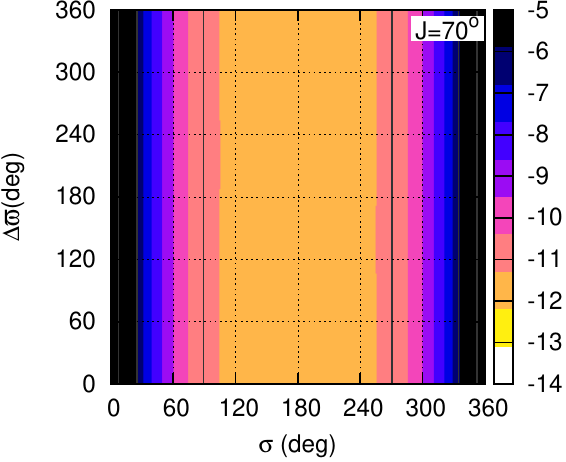}}\\
\mbox{\includegraphics[width=0.51\columnwidth]{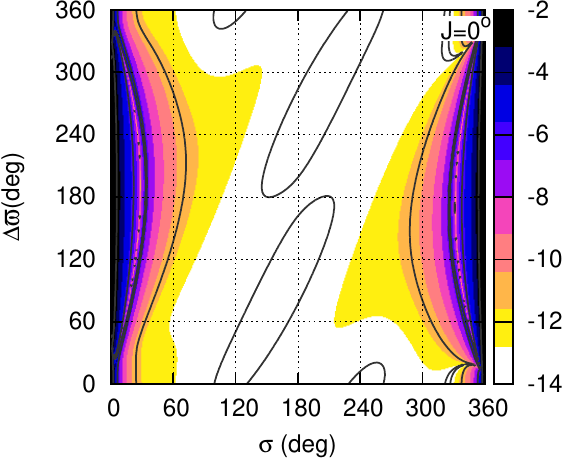}} \mbox{\includegraphics[width=0.51\columnwidth]{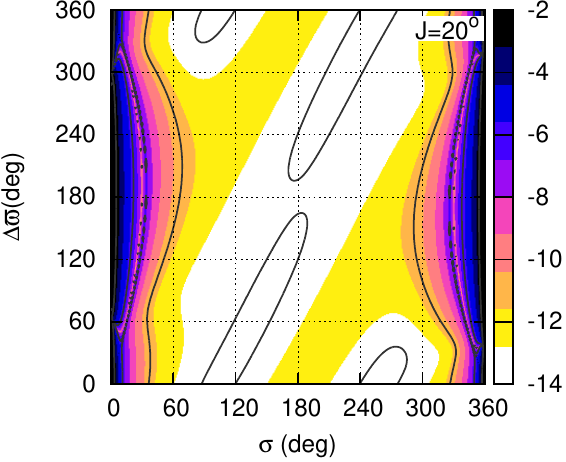}} \mbox{\includegraphics[width=0.51\columnwidth]{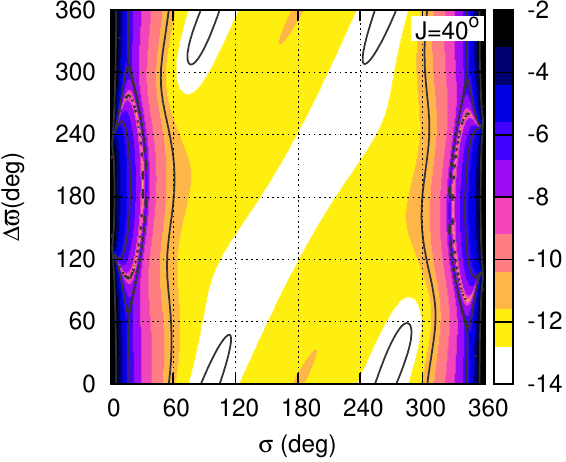}} \mbox{\includegraphics[width=0.51\columnwidth]{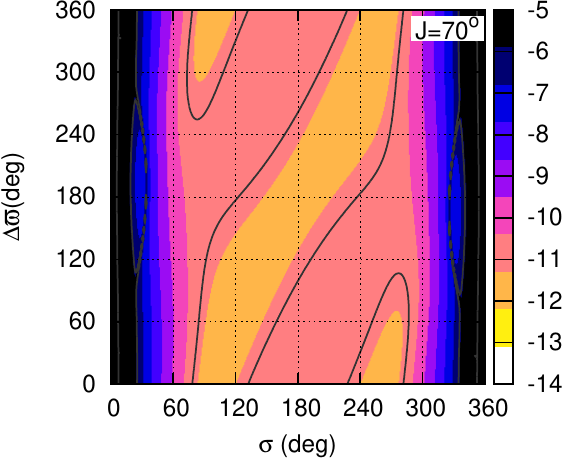}}\\
\caption{The relative error between the semianalytical averaged Hamiltonian, $\bar{\mathcal{H}}$, and the analytical expansion, $\mathcal{H}_2$. We consider two Jupiter-like planets at 1 au with $e_i=0.01$ (top row), and $e_i=0.15$ (bottom row). White regions are the most adequate to moderate the dynamics using the $\mathcal{H}_2$ model.}
 \label{figQS}
\end{figure*}

\begin{figure*}
\centering
 \mbox{\includegraphics[width=0.68\columnwidth]{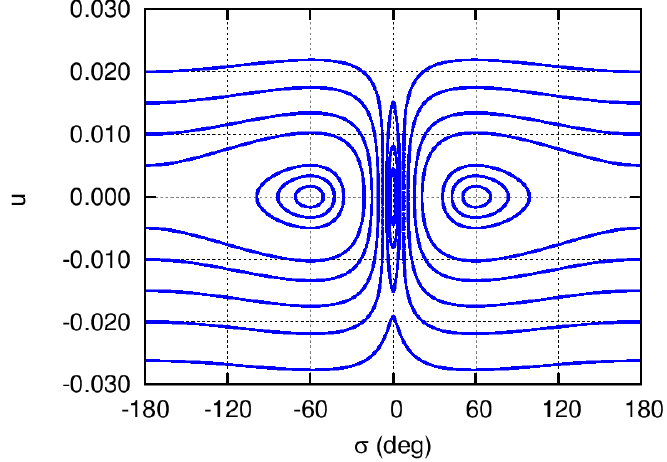}} \mbox{\includegraphics[width=0.68\columnwidth]{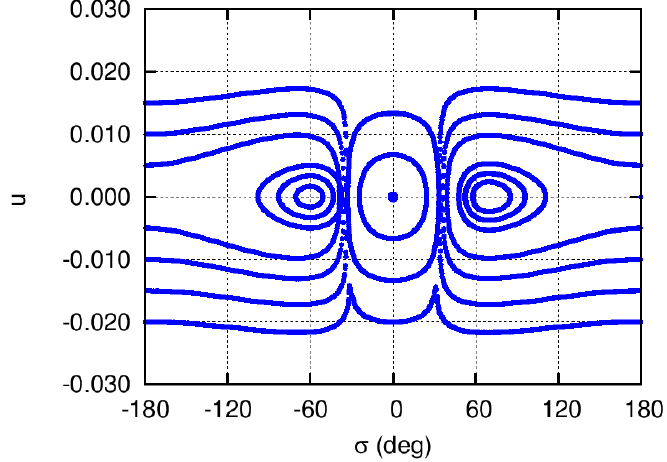}} \mbox{\includegraphics[width=0.68\columnwidth]{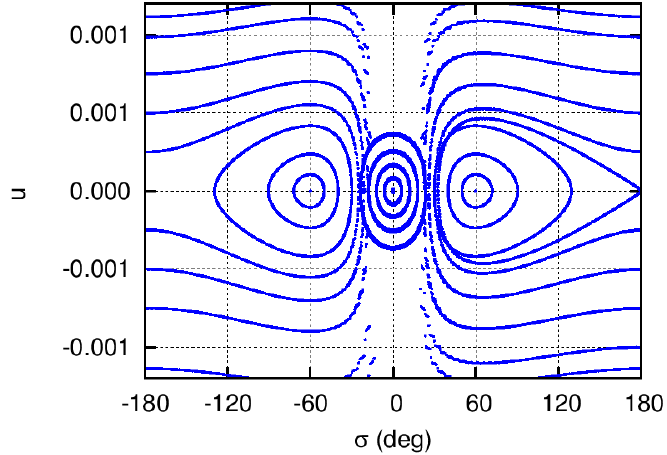}}
\caption{The phase space described by the semianalytical model $\bar{\mathcal{H}}$ in the plane ($u,\sigma$) for same initial conditions than Fig.\ref{fig1}. Left panel. Jupiter pair planets in quasi-circular orbits ($e_i=0.01$). Middle panel. Jupiter pair planets with moderate eccentricities ($e_i=0.15$). Right panel. Earth-like planets ($m_i=3\times10^{-6}M_\odot$) in quasi-circular orbits.}
 \label{fig-sem}
\end{figure*}

Since D'Alembert's relation provides a restriction for the $j_{k}$ coefficients, $\sum_{k} j_{k} = 0$, $t_2$ does not appear in $\varphi$ ($t_2$ is a cyclic angle). As a consequence, the associated action $T_2$ is a constant of motion and we can reduce our problem by one degree of freedom. Hence, our election of canonical variables leads to $T_2=\frac{1}{2}(H_1+H_2)=\frac{1}{2}{\cal AM}$ (half the orbital Angular Momentum of the system). 

Then, the Hamiltonian function can be expressed as $\mathcal{H} = \mathcal{H}_0 + \mathcal{H}_1$, where $\mathcal{H}_0$ corresponds to the two-body contribution, 
\bea
\mathcal{H}_0 = -{\frac{\mu_1^2 \beta_1^3} { 2L_1^2}}-{\frac{\mu_2^2 \beta_2^3} { 2L_2^2}}.
\eea
The second term, $\mathcal{H}_1$, is the disturbing function which can be written as:
\bea
\mathcal{H}_1=-\mathcal{G}m_1m_2 \frac{1}{\Delta} + \mathcal{T}_1, 
\eea
where $\Delta$ is the instantaneous distance between the two planets, and $\mathcal{T}_1$ is the indirect part of the potential energy of the gravitational interaction.

Obviously, the equations \ref{eq.poincare} and \ref{eq.coorbital} achieve the same results, but the latter has only 5-degrees-of-freedom, imposing the conservation of the angular momentum. 

The next step is average of the Hamiltonian over the fast angle $s_1$. This procedure can be performed numerically, allowing us to evaluate the averaged Hamiltonian $\bar{\mathcal{H}}$ as:
\be
\label{eq.aver}
\bar{\mathcal{H}}(R_1,R_2,S_2,T_1,\sigma,\Delta\varpi,s_2,t_1;{S_1},{\cal AM}) \equiv \frac{1}{4\pi} \int_0^{4\pi} \mathcal{H}\, d s_1. 
\ee
In the averaged variables, $S_1$ is a new integral of motion which, in analogy to other mean-motion resonances, we identify as the {\it scaling parameter, i.e. ${\cal K}$ }. 

$\bar{\mathcal{H}}$ constitutes a system with four degrees of freedom in the canonical variables ($R_1,R_2,S_2,T_1,\sigma,\Delta\varpi,s_2,t_1$), parametrized by the values of both ${\cal K}$ and ${\cal AM}$. Since the numerical integration depicted in equation (\ref{eq.aver}) is equivalent to a first-order average of the Hamiltonian function \citep[e.g][]{Ferraz-Mello_2007}, only those periodic terms with $j_1+j_2=0$ remain in $\bar{\mathcal{H}}$ (see Eq. \ref{Dal}). 

We have compared the semianalytical model averaged over the fast angle with the filtered N-body integrations. The filter was made using a low pass FIR digital filter (\citet{Carpino_etal_1987}) to eliminate all periodic variations with a period smaller than $3$ years. Needless to say, these results match better than those reproduced by the second order Hamiltonian, $\mathcal{H}_2$, but much slower. Due to the fact that we do not have restrictions for any configuration, $\bar{\mathcal{H}}$ is more adequate in the whole coorbital resonance. As an example, we show the results for an initial condition corresponding to a $QS$ orbit in Figure \ref{fig.inteQS}. No significant differences are appreciated for actions, angles, frequencies neither for orbital elements. 

Moreover, combining the information from Eq. \ref{eq.poincare} with the expansions in Eq. \ref{eq.expansion} we easily identify $S_1,S_2$ and $T_2$ as constants of motion. Thus, we can deduce the coupling in the orbital elements in the averaged models, namely
\begin{equation}
\begin{split}
      \beta_1\sqrt{\mu_1 a_1}+\beta_2\sqrt{\mu_2 a_2} &= \text{const} \\
       L_1 {e_1}^2+L_2 {e_2}^2 & \simeq \text{const} \\
       L_1 {e_1}^2 \cos (i_1) + L_2 {e_2}^2 \cos (i_2) &\simeq \text{const}
\end{split}
\label{eqs.coorb}
\end{equation}
From previous equations, the coupling between $e$ and $i$ present in the Lidov-Kozai resonance is not obvious (see Section \ref{3d}). 

We explore the parameter space ($\sigma, \Delta \varpi$) and plot the relative difference between the mean Hamiltonian, $\bar{\mathcal{H}}$, and the average $\mathcal{H}_2$ model. The $QS$ region\footnote{The region defined around ($\sigma,\Delta\varpi$)$=$($0^\circ,180^\circ$). See \citet{Giuppone_etal_2010} to identify the regions of motion within coorbital resonance.} shows more discrepancy, even considering Neptune-like planets in quasi circular orbits. {In the Figure \ref{figQS} we construct a colour map in the plane ($\sigma,\Delta\varpi$) considering two Jupiter-like planets with quasi circular initial conditions, $e_i=0.01$, and in eccentric orbits, $e_i=0.15$. Also, we identify the initial mutual inclination, $J$, in each panel. Outside the $QS$ region, the relative difference between Hamiltonians does not exceed $10^{-14}$, justifying the region of validity for the $\mathcal{H}_2$ model}. Furthermore, Analytical models valid for the $QS$ or ``eccentric retrograde satellite orbits'', were developed by \citet{Mikkola_etal_2006} and \citet{Sidorenko_2014}, but are only valid in the frame of the RTBP, considering small inclinations. 

To illustrate the validity of this semianalytical model Figure \ref{fig-sem} shows the integrations for the same initial conditions of Figure \ref{fig1}. Obviously, if the mutual inclination, $J$, or eccentricities, $e_i$, increases, the analytical Hamiltonian $\mathcal{H}_2$ is more inexact. The semianalytical model eliminates the short periodic terms and, is which is easier to identify the different types of motion. 

\section{Phase space in the 3D case}\label{3d}

Our intention in this section is to find the different types of stable orbits present in the inclined systems for the 1:1 MMR.

\citet{Voyatzis_etal_2014} studied systems that migrate under the influence of dissipative forces that mimic the effects of gas-driven (Type II) migration. They demonstrated that sometimes excitation of inclinations occurs during the initial stages of planetary migration. In these cases, {\em vertical critical orbits} may generate stable families of 3D periodic orbits, which drive the evolution of the migrating planets to non-coplanar motion. Their work focuses on the calculus of the vertical critical orbits of the 2:1 and 3:1 MMR, for several values of the planetary mass ratio. 
{In hierarchical systems, the secular resonance Lidov-Kozai (LK) provides conditions for periodic orbits for inclined systems, and its centre of libration is located at $\omega =\pm90^\circ$ \citep[e.g.][]{Lidov_1961, Kozai_1962,Kinoshita_Nakai_2007}}. The secular Hamiltonian of RTBP (expanded up to quadrupole order in the semi-major axis ratio $a_1/a_2$ and averaged with respect to the fast periods $\lambda_1$ and $\lambda_2$) does not depend on $\Omega$. Hence, its conjugated action is a constant; consequently,
\begin{equation}
\sqrt{{\cal G}\,m_{0}\,a(1-e^2)}\,\cos(i)=const \ .
\label{eqkozai} 
\end{equation}

Evidence of Lidov-Kozai resonance for planetary systems was found in the 2:1 MMR \citep{Antoniadou_etal_2013} and compared with the circular RTBP. As was pointed by \citet{Libert_tsiganis_2009} the stability of some inclined exoplanetary systems may be associated with the LK resonance. Moreover, \citet{Morais_2016} showed that the LK resonance is present for retrograde orbits as well as in prograde orbits and plays a key role in coorbital resonance capture for circular RTBP.  

{The LK resonance occurs in hierarchical planetary systems and can be identify dynamically. The centre of this resonance occurs when the mutual inclination between the bodies and the shape of their orbits remain \textit{frozen} in the integration. This fact occurs at $\Delta\varpi =\pm 90^\circ$. Thus, we identify the centre of LK resonance throughout different dynamical maps when the amplitude of oscillation for $e$, $J$ and $\omega$ tends to zero.}

In our development we average over the sum $\lambda_1+\lambda_2$ instead of $\lambda_1$, $\lambda_2$, obtaining new conserved quantities. Nonetheless, at the limit when the mass ratio goes to zero ($m_2/m_1 \rightarrow 0$) we recovered the results from RTBP, from conservation of angular momentum (see Eq. \ref{eq.coorbital} and Eq. \ref{eqkozai}).

\begin{figure}
 \centering
\mbox{\includegraphics[width=0.44\columnwidth]{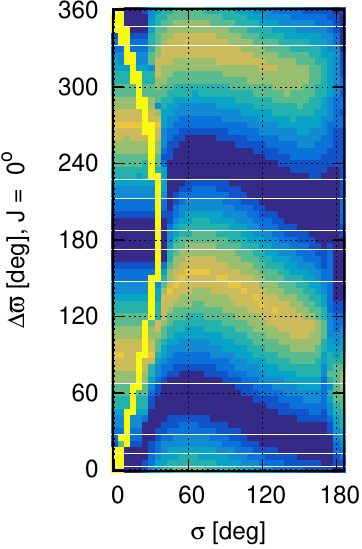}} \mbox{\includegraphics[width=0.484\columnwidth]{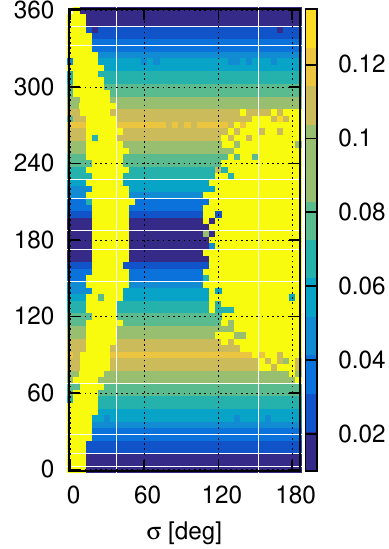}}\\
\mbox{\includegraphics[width=0.44\columnwidth]{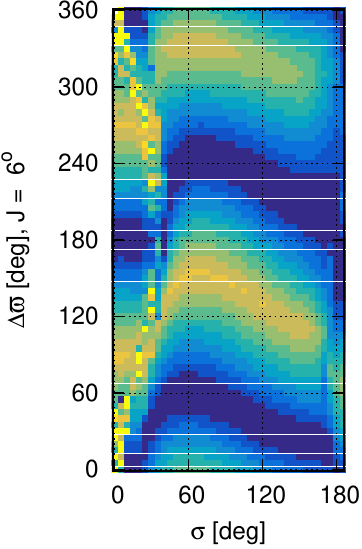}} \mbox{\includegraphics[width=0.484\columnwidth]{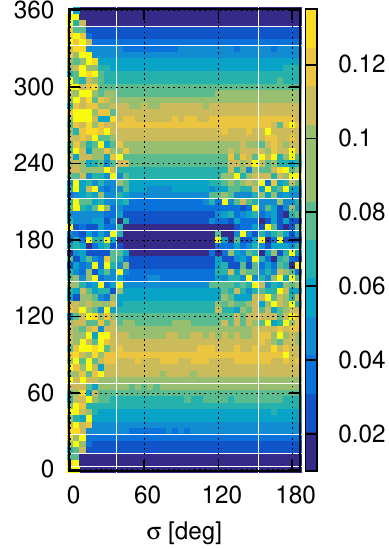}}\\
\mbox{\includegraphics[width=0.44\columnwidth]{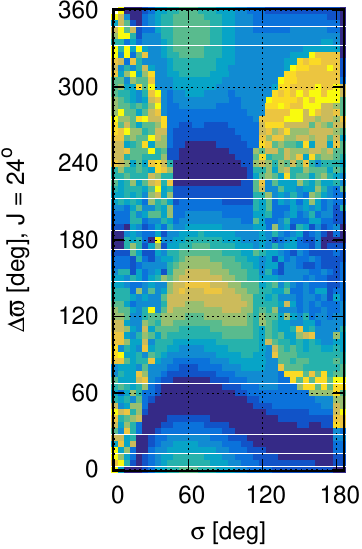}} \mbox{\includegraphics[width=0.484\columnwidth]{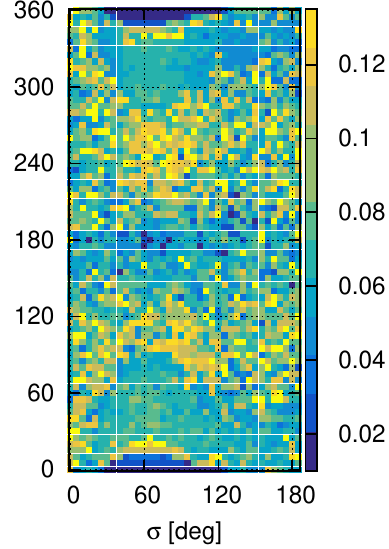}}\\
\mbox{\includegraphics[width=0.44\columnwidth]{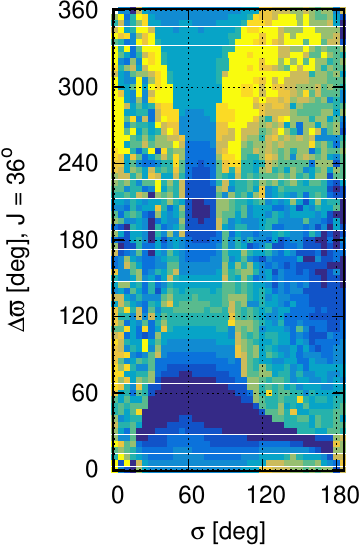}} \mbox{\includegraphics[width=0.484\columnwidth]{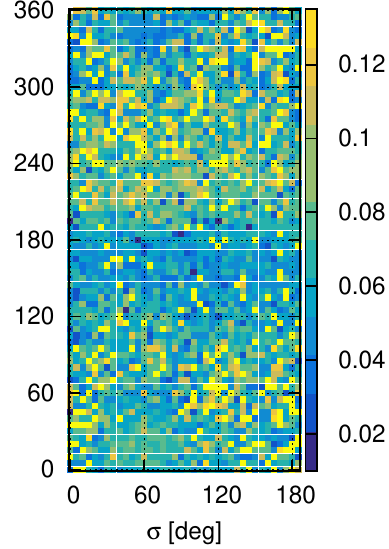}}\\
\caption{Initial conditions integrated for $10^4$ periods with $e_1=e_2$ and $m_1=m_2=4 M_\oplus$. Initial conditions for left column have $a_1=a_2=1$ au, while for right column $a_1$=1.004838 and $a_2$=0.99517 au. Colour scale represents the amplitude variation of $e_2$. Each row correspond to a different initial $J$.}
 \label{figJ}
\end{figure}

\begin{figure*}
 \centering
\mbox{\includegraphics[width=18 cm]{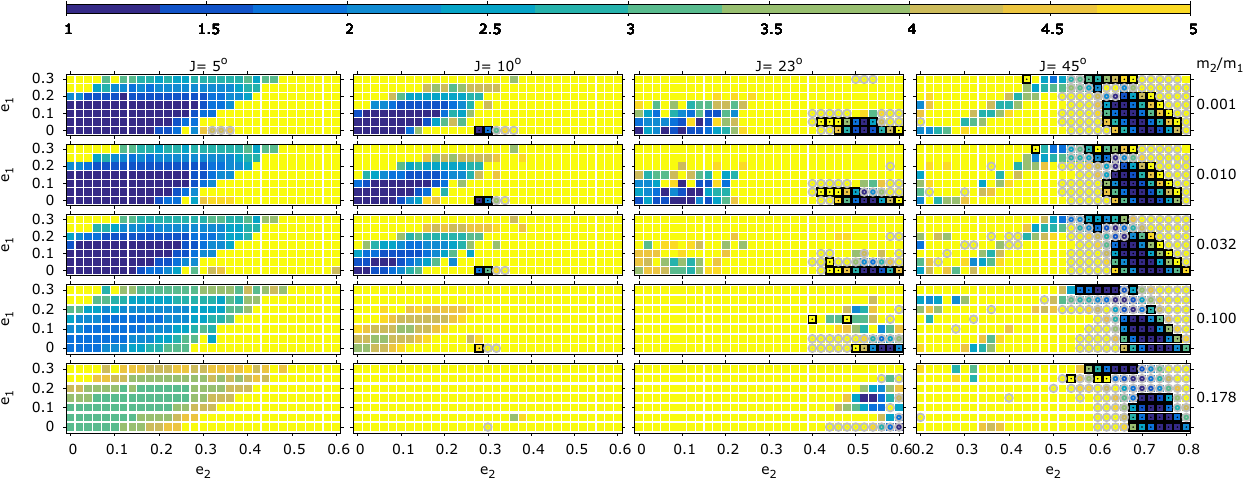}}
\vspace{-1.5em}
\caption{Initial conditions integrated for $10^5$ periods. Black squares correspond to amplitude of $\omega_2<10^\circ$, grey circles for $10^\circ<\omega_2<20^\circ$. For the remaining initial conditions $\omega_2$ circulates very slowly. Colour scale is proportional to the oscillation variation of $J$.}
\vspace{-2em}
 \label{figkoz-e1e2}
\end{figure*}

\begin{figure*}
 \centering
\mbox{\includegraphics[width=0.67\columnwidth]{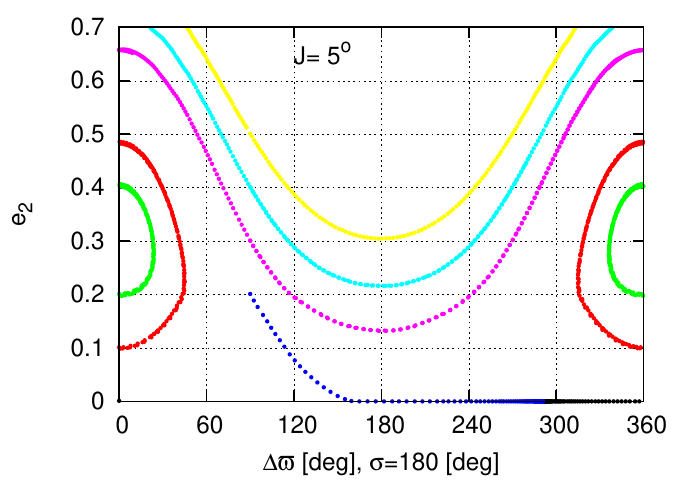}}  \mbox{\includegraphics[width=0.67\columnwidth]{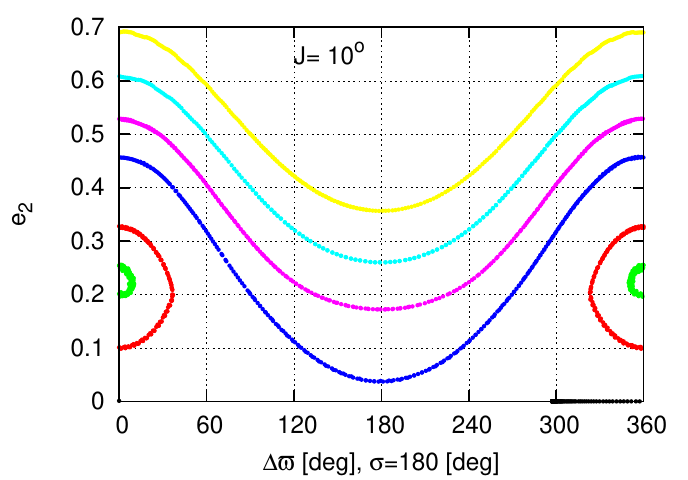}}  \mbox{\includegraphics[width=0.67\columnwidth]{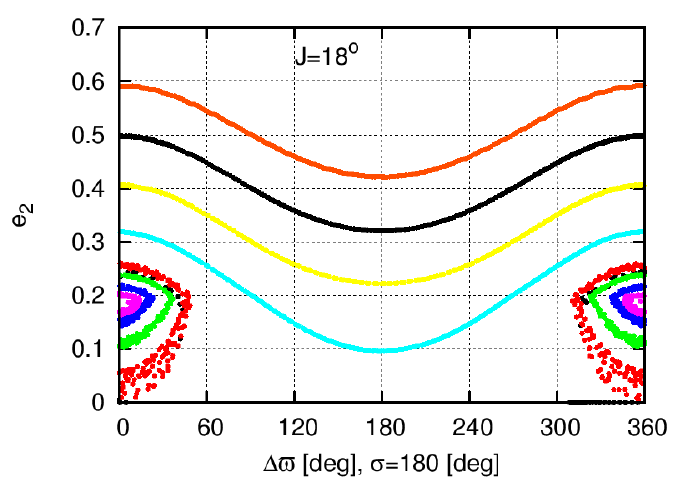}}  \\
\mbox{\includegraphics[width=0.67\columnwidth]{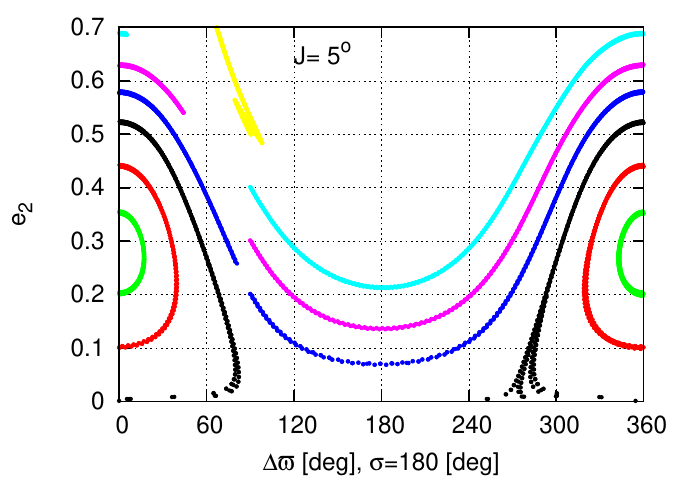}}    \mbox{\includegraphics[width=0.67\columnwidth]{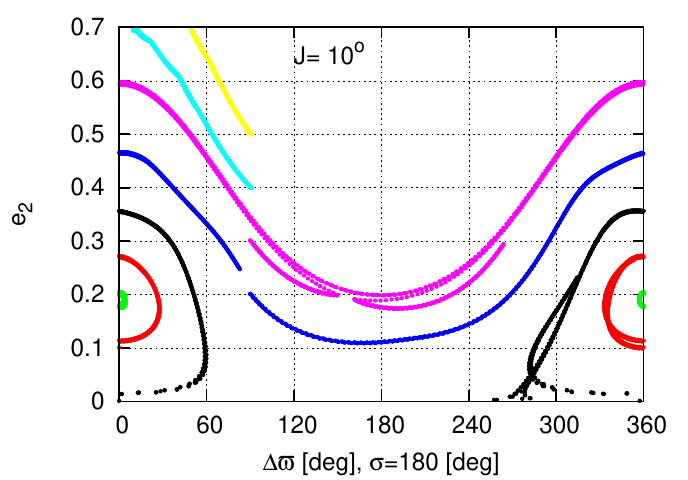}}    \mbox{\includegraphics[width=0.67\columnwidth]{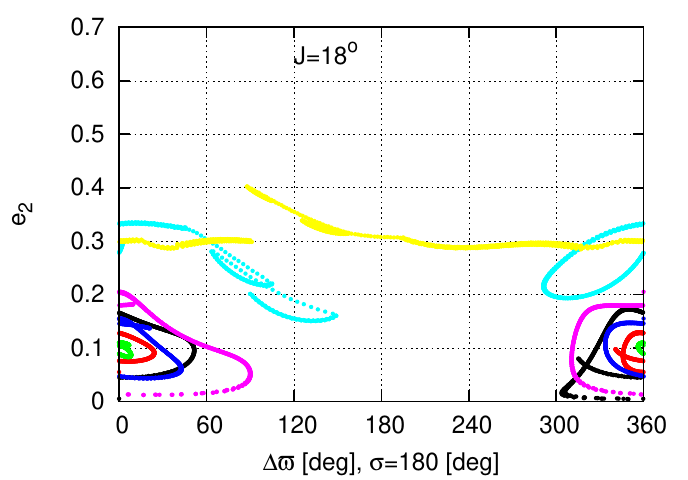}}     \\
\mbox{\includegraphics[width=0.67\columnwidth]{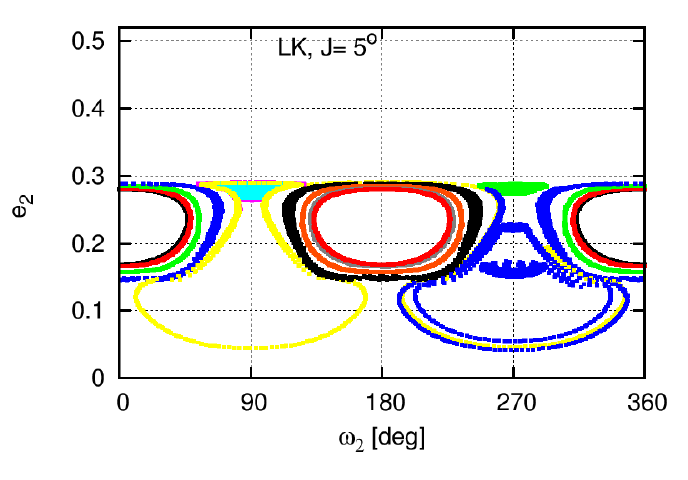}}         \mbox{\includegraphics[width=0.67\columnwidth]{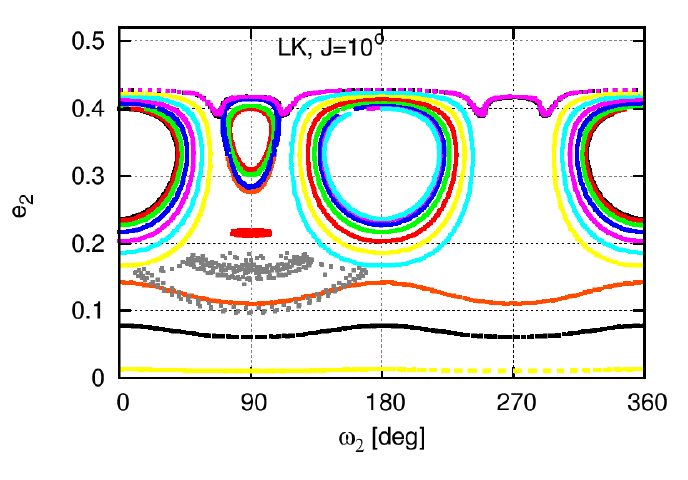}}         \mbox{\includegraphics[width=0.67\columnwidth]{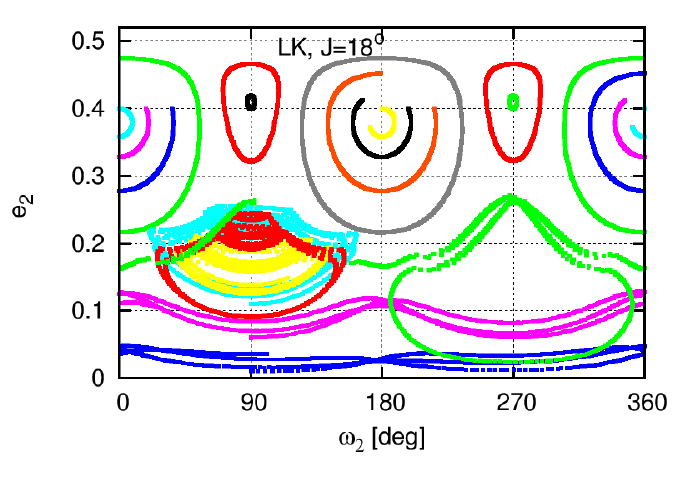}}   \\
\vspace{-1em}
 \caption{The phase space for a given value of $\mathcal{AM}$ and different initial mutual inclinations ($J=5^\circ$, $J=10^\circ$, $J=18^\circ$ from left to right). First and second row correspond to phase space using $\mathcal{H}_{2}$ and N-body integrations for $3\times 10^5$ periods respectively in the plane ($\Delta\varpi,e_2$) while the bottom row correspond to integrations on the plane ($\omega_2,e_2$) with $m_1=3 \times 10^{-6} M_\odot$, $m_2=3 \times 10^{-9} M_\odot$. Each colour represents the evolution of a different initial condition.}
 \label{figkoz-rtbp}
\end{figure*}

{The Figure \ref{figJ} shows the variation of oscillation for $e_2$ in the plane ($\sigma,\Delta\varpi$), setting two equal mass planets at low eccentric orbits ($e_1=e_2=0.15$) for several different values of initial mutual inclinations. At the left panel, where $a_1=a_2$, when initial mutual inclination is low we can identify QS orbits at ($0^\circ,180^\circ$), $L_4$ at  ($60^\circ,60^\circ$), $AL_4$ at ($\simeq 70^\circ,\simeq 250^\circ$). As the initial mutual inclination increases, the regions of periodic orbits shrink being only robust the $L_4$ condition that survives even for $J=36^\circ$ (tiny dark region at the bottom panel).
We have realized that setting $a_1=a_2$ does not give any further information about the possible existence of HS neither the Lidov Kozai resonance. Thus, following our results from Section \ref{model} we set $u=1.2 U_3$ ($a_1$=1.004838 and $a_2$=0.99517), showing the results in the right column of Figure \ref{figJ}. There, the $HS$ region appears at $\Delta\varpi = 0^\circ,180^\circ$, however it is not present for $J=0^\circ$ at ($\sigma,\Delta\varpi$)=($\simeq 180^\circ,\simeq 180^\circ$)\footnote{Besides, is present for $J=0^\circ$ with $a_1=a_2$. Its Megno value shows that is highly chaotic.}.}

{To optimize the identification of the region where the LK configuration appears, we study the plane $(e_1,e_2)$ and initial conditions with $\omega_2=90^\circ$, $\sigma=180^\circ$, $t_1=180^\circ$, $\Delta\varpi=s_1=s_2=0^\circ$, and $u=1.2 U_3$. In this plane we varied $J$ for different mass ratios, ranging from $\log(m_2/m_1)=-3$ to $0$. Figure \ref{figkoz-e1e2} resumes the results with a colour scale proportional to the variation of the mutual inclination, $J$. We identify two regions of periodic orbits. One region corresponding to $e_1 \simeq e_2$, which is easier to identify for low mutual inclination (see first column, $J=10^\circ$) and the other one corresponding to $e_1 \simeq 0$ and $e_2>0.3$ depending on $J$ and the mass ratio, that we refer as LK region. For a very low mass ratio, $m_2/m_1 \simeq 0.001$ (near RTBP conditions, top row in the figure), it is easy to find the LK resonance in the range of mutual inclination $10^\circ<J<50^\circ$. This configuration is only found up to $m_2/m_1 \simeq 0.178$ for very high values of $e_2$.}

{In Figure \ref{figkoz-rtbp} we plot both regions in the parameter space setting $m_2/m_1 \simeq 0.001$.  Top and Middle rows shows integrations with the $\mathcal{H}_2$ model and the N-body code respectively, in the HS region on the plane ($\Delta\varpi,e_2$). We used the same colours in both panels to facilitate the comparison between them. Evidently, the $\mathcal{H}_2$ model is limited to small (or even moderate) inclinations and eccentricities, reproducing very well the parameter space with the oscillation centres slightly displaced. We numerically verified that the results are indistinguishable between using $m_2=0$ (RTBP) or $m_2/m_1=10^{-3}$. Although, the systems are well reproduced for moderate mutual inclinations, the interactions between the bodies are evident for some orbits showing chaotic motion in the N-body integrations.}  

\begin{figure}
 \centering
\mbox{\includegraphics[width=\columnwidth]{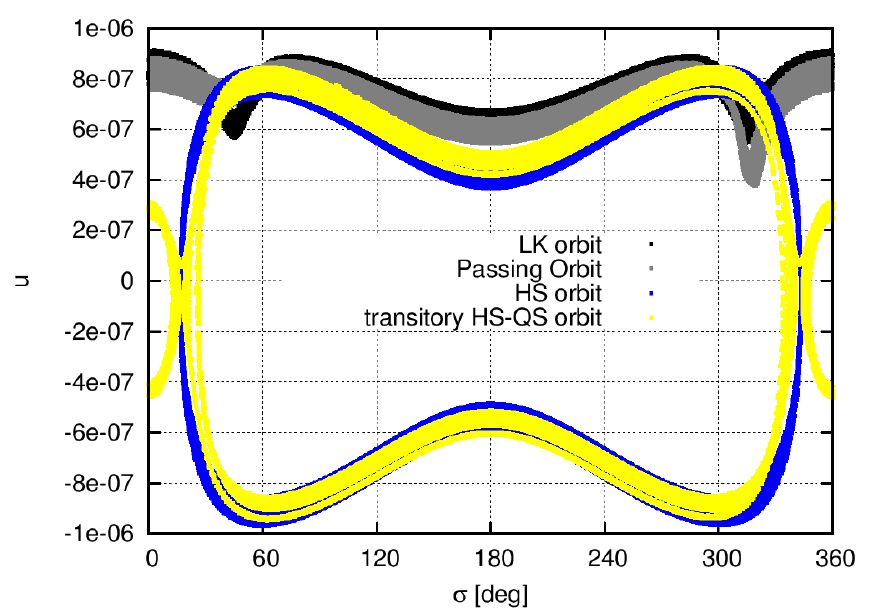}}
\vspace{-1.5em}
\caption{Examples of dynamical regimes present in Figure \ref{figkoz-rtbp}. See text for more detail.}
\vspace{-2em}
 \label{LK-regimes}
\end{figure}

{On the other hand, at the bottom row of Figure \ref{figkoz-rtbp} we show the LK region on the plane ($\omega_2,e_2$) for the same values of initial $J$. The results for LK at $J=18^\circ$ agree with \citet{Namouni_1999} for the RTBP inside the 1:1 MMR. The LK region is a mixture of dynamical regimes and it was insightful depicted numerically by \citet{Namouni_1999}. In Figure \ref{LK-regimes} we show these different kind of motions in the plane ($u,\sigma$) for $J=18^\circ$ using the same colours and conditions that in bottom right hand-panel of Figure \ref{figkoz-rtbp}. In the region of low eccentricities the motion is of horseshoe-type and $\omega_2$ circulates. Regions at $\omega_2$ = $0^\circ$ or $180^\circ$, where $\omega_2$ librates with moderate values of $e_2$, are those corresponding to passing orbits. We can identify the LK resonance at $\omega_2 =\pm 90^\circ$ with $e_2 \simeq 0.4$ (where $\omega_2$ librates), and near to the LK resonances is the vase-like domain where transitions between $HS-QS$ orbits are present. However, the analytical $\mathcal{H}_2$ model was not able to reproduce the structure of the phase space. Also, the phase space for $J=5^\circ$ only shows transition orbits and temporary $HS-QS$ orbits.} 

\begin{figure}
 \centering
\mbox{\includegraphics[width=\columnwidth]{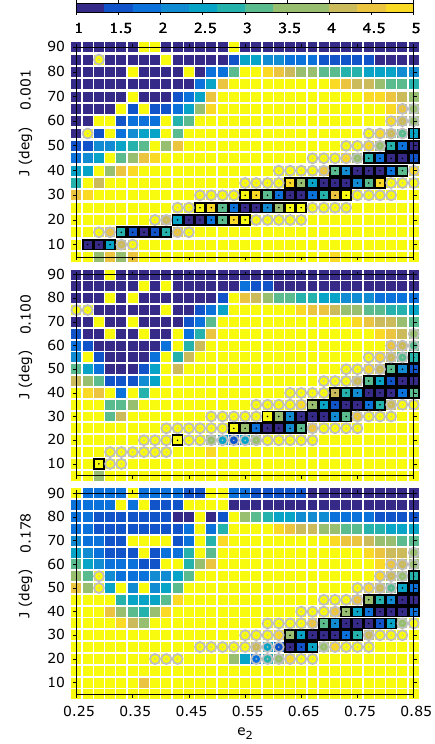}}
\vspace{-1.5em}
\caption{Location of LK resonance centre of depending on mass ratio and initial mutual inclination. The more massive planet is on circular orbit $e_1=0$ with remaining orbital elements as in Figure \ref{figkoz-e1e2}. The initial conditions were integrated for $10^5$ periods. Black squares correspond to amplitude of $\omega_2<10^\circ$, grey circles to $10^\circ<\omega_2<20^\circ$. For the remaining initial conditions $\omega_2$ circulates very slowly.}
\vspace{-2em}
 \label{fig-Dj}
\end{figure}

\begin{figure}
 \centering
\mbox{\includegraphics[width=\columnwidth]{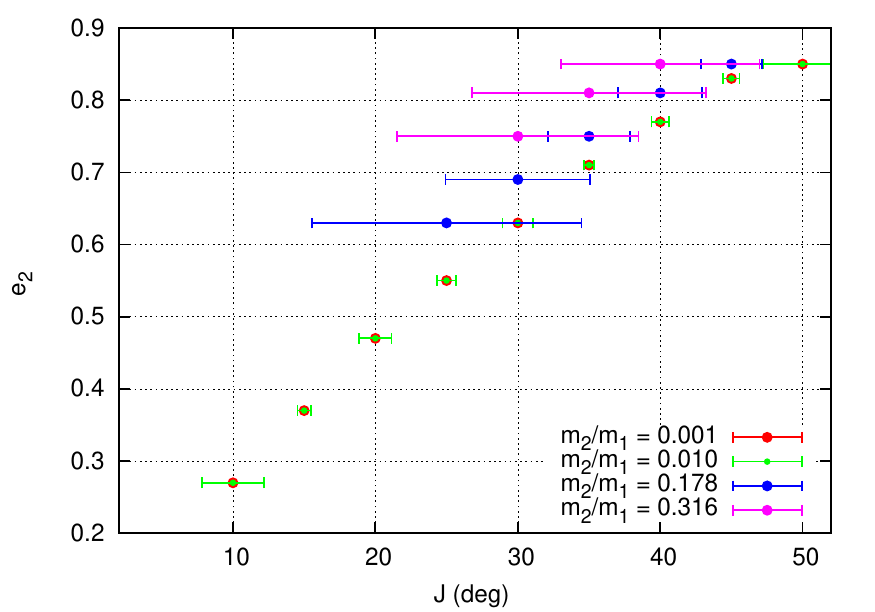}}
\vspace{-1.5em}
\caption{Location of the LK resonance centre depending on the mass ratio and the initial mutual inclination. Error bars are proportional to oscillation of $\omega_2$.}
\vspace{-2em}
 \label{fig-fam}
\end{figure}

To resume the location of the LK resonance, we plot in the plane ($e_2,J$) the amplitude of oscillation of $J$, setting $e_1=0$ for different mass ratios (see Figure \ref{fig-Dj}). We can perfectly identify that the LK resonance is present up to $J\simeq50^\circ$ and its appearance strongly depends on the mass ratio. Thus, for example, earth-like planets can be in the centre of the LK resonance at low or high inclinations. In Figure \ref{fig-fam} we only plot the points corresponding to the minimum amplitude of oscillation of $J$ for several mass ratios. When $m_2/m_1 \rightarrow$ 0.3, the LK resonance almost dissipates and the strong interactions cause that the amplitude of oscillation for $\omega_2$ increases. However, they are regular orbits, according to their Megno value ${<}Y{>} (\lesssim 2.02$).

\begin{figure*}
 \centering
\mbox{\includegraphics[width=0.68\columnwidth]{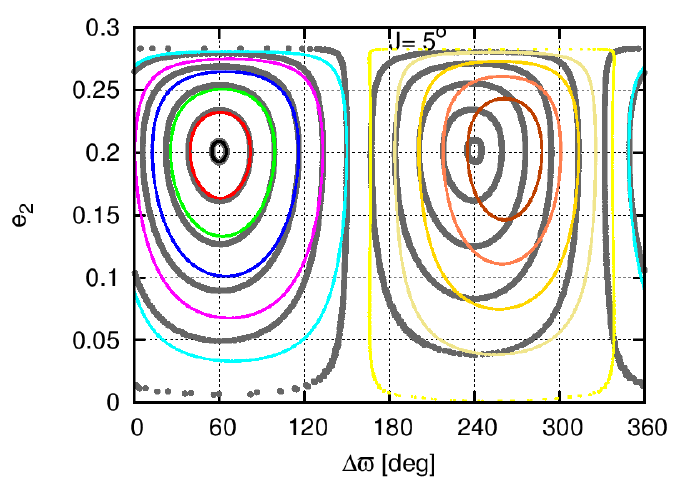}}    \mbox{\includegraphics[width=0.68\columnwidth]{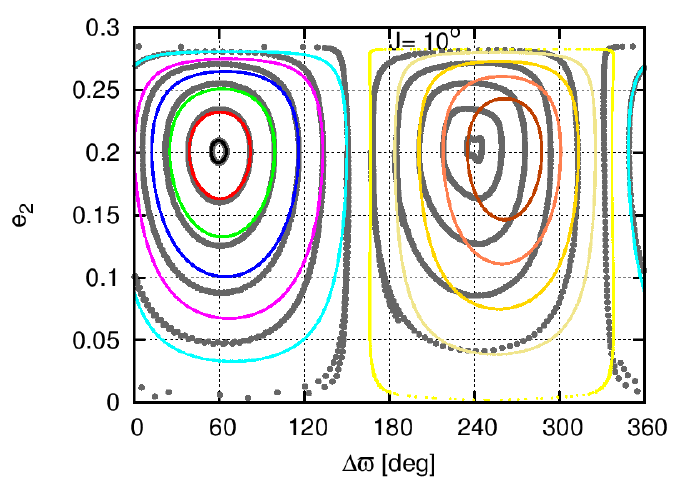}}  \mbox{\includegraphics[width=0.68\columnwidth]{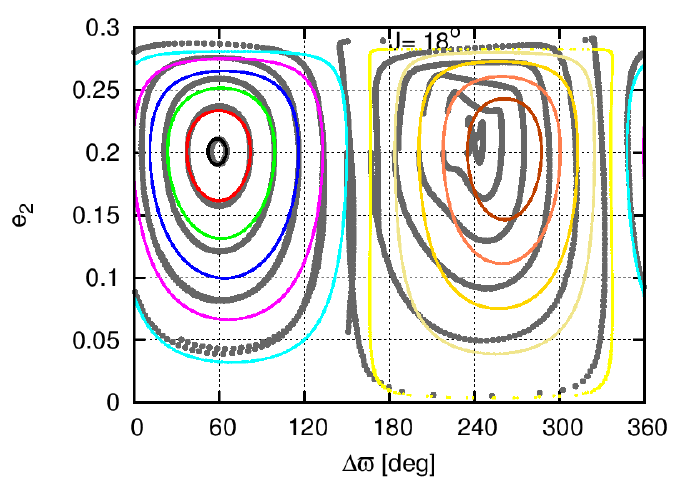}}     \\
\mbox{\includegraphics[width=0.68\columnwidth]{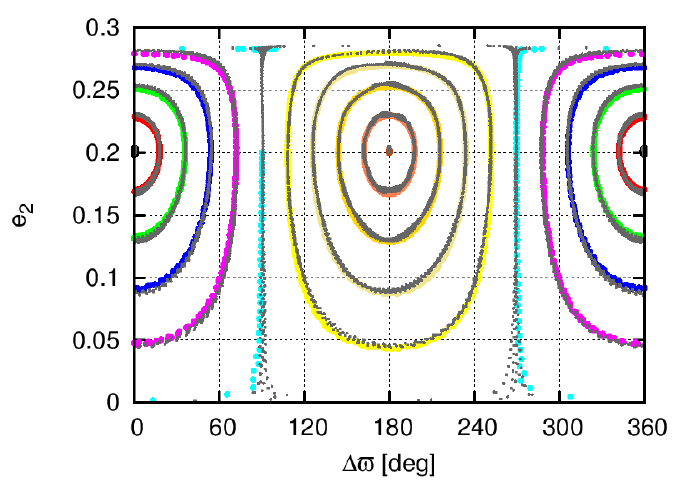}}      \mbox{\includegraphics[width=0.68\columnwidth]{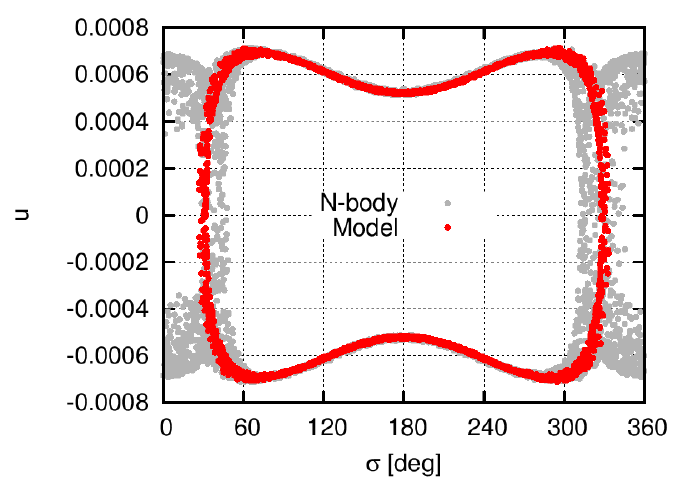}}    \mbox{\includegraphics[width=0.68\columnwidth]{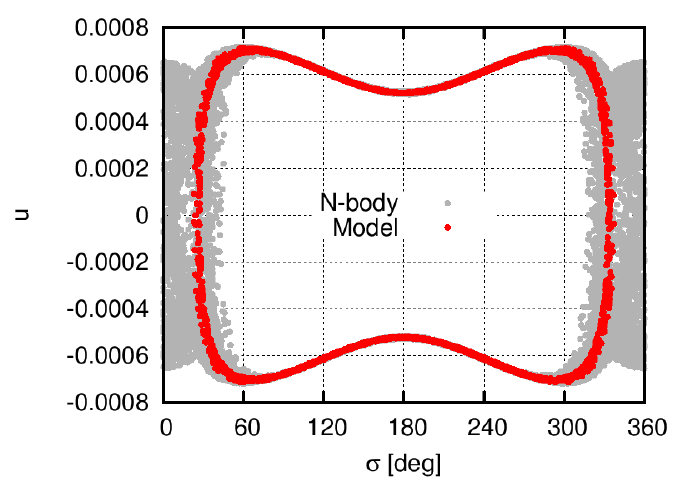}}   \\
 \caption{The phase space for a given value of $\mathcal{AM}$ and initial mutual inclination $J=5^\circ$, $J=10^\circ$, $J=18^\circ$ (from left to right respectively). Top row. Two Jupiter-like planets with $a_1=a_2=1$, $\sigma=60^\circ$ (Lagrangian region). Bottom row. Two planets in the HS region, with mass 3 $\times 10^{-6} M_\odot$, $\sigma=180^\circ$, and $u=1.2 U_3$. We plot with grey dots the N-body integrations, and with colours the $\mathcal{H}_2$ model integrations.}
 \label{figkoz-gtbp}
\end{figure*}

\begin{figure*}
 \centering
\mbox{\includegraphics[width=0.68\columnwidth]{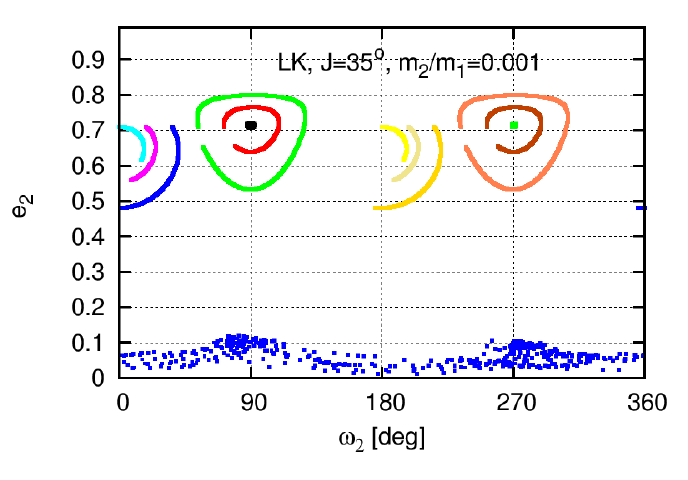}}    \mbox{\includegraphics[width=0.68\columnwidth]{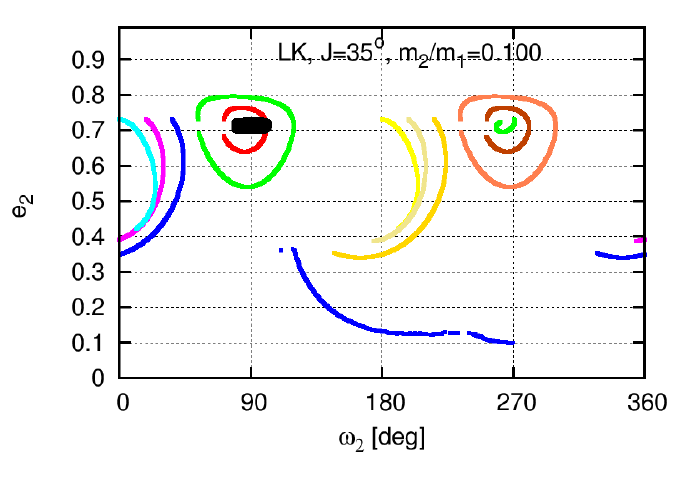}}  \mbox{\includegraphics[width=0.68\columnwidth]{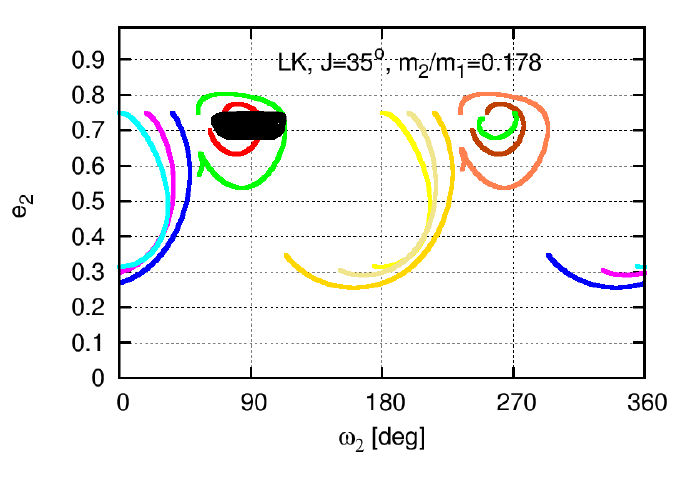}}     \\
\caption{The phase space for a given value of $\mathcal{AM}$ to show the LK resonance centre for $J=35^\circ$ for three different mass ratios.}
 \label{figkoz-35}
\end{figure*}

Contrary to the case of Figure \ref{figkoz-rtbp}, the case of general problem (both planets with similar masses) is slightly different, and easier to analyse on the plane ($\Delta\varpi$,$e_2$). The averaged analytical $\mathcal{H}_2$ model works well even for high inclinations and we are able to analyse the structure depicted by numerical integrations. The Figure \ref{figkoz-gtbp} shows the phase space for three different mutual inclinations, $J$, and moderate eccentricities (initially, $a_1=a_2=1$ ua, $m_1=m_2=3 \times 10^{-6} M_\odot$, and $e_1=e_2=0.2$). When the initial conditions have $\sigma=60^\circ$ (top row) we can easily identify two islands of stability corresponding to $L_4$ (at $\Delta\varpi=60^\circ$) and $AL_4$ (at $\Delta\varpi\simeq 240^\circ$) configurations. The $L_4$ region is very well depicted by the model, even for $J=18^\circ$, but the $AL_4$ region shifts artificially the centre for $\Delta\varpi \rightarrow 270^\circ$; besides the amplitude of $e_2$ is well represented. This effect is due to the limitations of expansions. For quasi circular orbits this shift vanished. 

In bottom row of Figure \ref{figkoz-gtbp}, when $\sigma=180^\circ$, we set $u=1.2 U_3$. The oscillation centres are around $\Delta\varpi\simeq 0^\circ$ and $\Delta\varpi\simeq 180^\circ$. Indeed, the central point with $\Delta\varpi \simeq 180^\circ$ may correspond to the Euler configuration $L_3$ ($J\rightarrow 0^\circ$), which is unstable in the RTBP. When $J$ increases, this family is the only one that survives, although it seems chaotic in this plane. The other centre, around $\Delta\varpi\simeq 0^\circ$, correspond to the family identified as unstable by \citet{Hadjidemetriou_etal_2009, Hadjidemetriou_etal_2011} in the frame of coplanar planetary problem, using Jupiter planets. For $J \lesssim 20^\circ$ the model perfectly matches N-body integration. After that, this region become unstable. Then,  we  plot results in the projected plane ($\sigma, u$) for $J=10^\circ$ and $J=18^\circ$ to show that chaotic orbits in this region corresponds to $HS-QS$ transition orbits (N-body integration with grey points). However, the general behaviour of $HS$ is captured by the $\mathcal{H}_2$ model.

We tested a wide variety of systems with equal mass planets, from two Earth-mass planets to two Jupiter-mass planets with mutual inclinations as high as $60^\circ$ in the region previously identify as LK. We not find evidence of LK resonance. For high mutual inclinations, the systems are indeed strongly chaotic, and regular motion is allowed very close to the exact location of $L_4$ or $AL_4$, being the region around $L_4$ broader. 

Also, it is important to remark that as the mutual inclination is greater than 10$^\circ$, the systems exhibit chaotic behaviour if the initial conditions do not corresponds to the equilibrium solution ($e_1=e_2$, for $m_1=m_2=1 M_\oplus$). Nevertheless, we find some systems initially located at high mutual inclinations that can be in coplanar orbits after scattering.

{N-body integrations of Figure \ref{figkoz-35} shows the example of LK phase portraits in the plane ($\omega_2,e_2$) for $J=35^\circ$ and different mass ratios, with $m_1=1 M_\oplus$ and $e_1=0.001$. Note the centre of LK resonance located at $\omega_2=\pm90^\circ$. Low eccentric regime, $e_2 \lesssim 0.2$, is usually chaotic for this value of $J$. We use different colours to identify the evolution of initial conditions integrated for $3 \times 10^5$ periods, while the condition corresponding to $\omega_2$ = $90^\circ$ was integrated during $10^7$ periods. Is easy to see the importance of the forced oscillation around LK centre when $m_2/m_1 \rightarrow 0.2$, justifying the error bars in the Fig. \ref{fig-fam}}.

\begin{figure}
 \centering
\mbox{\includegraphics[width=8.0cm]{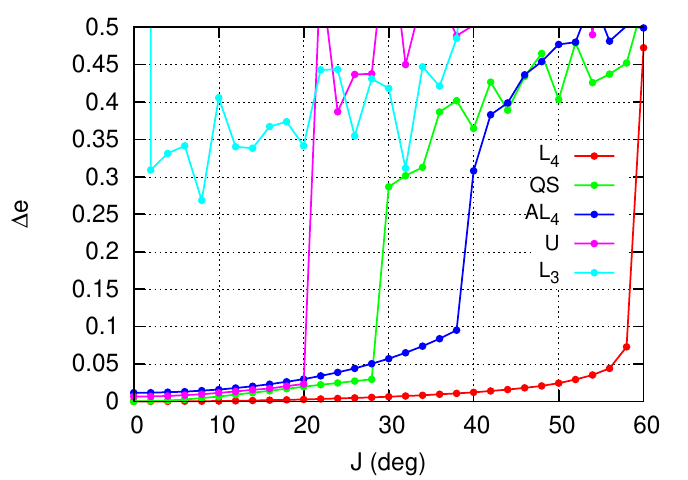}} \\
\mbox{\includegraphics[width=8.0cm]{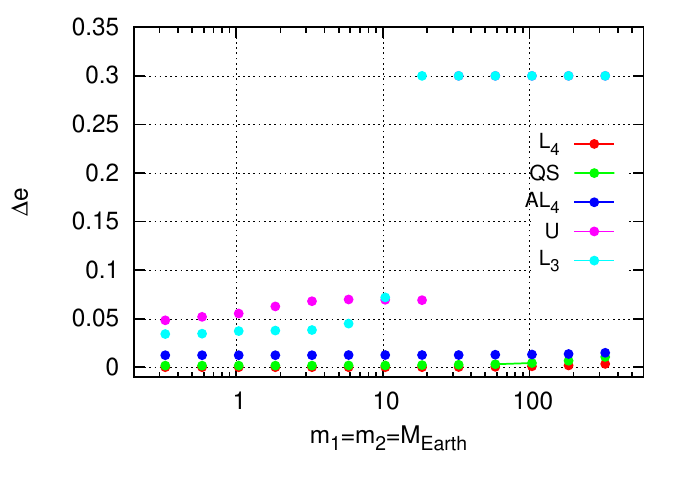}} \\
 \caption{Analysis of periodic orbits $L_4$, $AL_4$, $QS$, $L_3$, $U$ in the non-coplanar case using as indicator the variation of $\Delta e_2$. Results correspond to N-body integrations for $10^6$ periods. Top panel. Results considering two planets with masses $m_1=m_2=4 M_\oplus$, and $e_i=0.15$. {The smooth variation of $\Delta e_2$, when we increase $J$, is a good indicator of regular orbits (${<}Y{>} \simeq 2 $)}, although almost all the orbits survives at least for $10^6$ periods. Bottom panel. Analysis of stability for inclined systems depending on their masses (see text for more detail).}
 \label{fig-i}
 \vspace{-1.5em}
\end{figure}

{Finally we analyse the 3D configurations for periodic orbits mentioned in Figure \ref{Fig.schema}: $L_4$, $AL_4$, $QS$, $L_3$, $U$. {To construct the families of periodic orbits in the spatial case we began from the previous known results in the planar case and varied the mutual inclinations, $J$. For each family we checked that $\dot\sigma$=$\Delta\dot\varpi$=0, setting the remaining angles equal to zero. We believe that, this is a natural extension from the periodic orbits in the equal-mass planar case, although a more rigorous search should use the local extrema of the semianalytical Hamiltonian}. The top panel in Figure \ref{fig-i} shows the variation of amplitude of oscillation for $e_2$, $\Delta e_2$, for systems with different initial mutual inclinations and integrated during $10^6$ periods. For $L_4$, $AL_4$, and $QS$ orbits we set initial semi major axes $a_i=1$ ua, while for $L_3$ and $U$ we set $a_i$ using $u=1.2 U_3$. We calculate the Megno value for every orbit, ${<}Y{>}$, but we choose to show $\Delta e_2$ indicator because is easier to see the smooth \textit{degradation} of orbits as $J$ increases; alternatively $\Delta J$ is a good indicator too. The most regular orbits are those corresponding to $L_4$ configurations (even for $J \simeq 60\de$). $AL_4$ orbits are regular when $J\lesssim38\de$ and $QS$ orbits are regular up to $J\lesssim28\de$. {On the other hand, $U$-type configurations remain stable and bounded for choosen planetary masses (4 $M_\oplus$) when $J \lesssim 20\de$, although the evolution of orbital elements shows a slow chaos difussion}. The $L_3$ orbits are also interesting. For $J=0$ the orbits are unstable, yielding to close encounters between the planets, however for $J>0$ the orbits become stable for at least $10^6$ periods (${<}Y{>} > 5$). Even for $J\simeq60\de$ the orbits oscillate around $\Delta\varpi=180\de$, although when $0\de<J<20\de$ their chaoticity is more bounded. 

The bottom panel of Fig. \ref{fig-i} shows $\Delta e_2$, attained during the integrations, for several mass values of pair of planets. We choose to show $J=5\de$ to illustrate the general behaviour of the families. $L_4$, $AL_4$, and $QS$ configurations are regular and robust configurations in the range $0.3 M_\oplus < m_i < 1 M_{Jup}$. {The $U$-type orbits seems to be regular for masses $m_i \lesssim 10 m_\oplus$, despite long-term diffusion is observed and they remain in this configuration at least for 1 Gy; in contrast for systems with more massive planets ($m_i\gtrsim 20 m_\oplus$) close encounters causes expulsion of one planet ($a_i>2$ au) in less than $10^4$ periods. This same limit was observed for the coplanar case}. The $L_3$ configurations are chaotic but bounded for $m_i \lesssim 15 M_\oplus$ and, also like $U$ configurations, after this value of masses the systems are quickly destroyed. }

\vspace{-1em}
\section{Conclusions}\label{Conclusions}
We studied the three-body problem in the context of coorbital resonance, considering coplanar and spacial configurations. We followed several approaches: analytical ($\mathcal{H}$), averaged analytical ($\mathcal{H}_2$, Eqs. \ref{eq.expansion}), and semi-analytical ($\bar{\mathcal{H}}$, Eqs. \ref{eq.aver}). We found appropriate angles and actions (Eqs. \ref{eq.coorbital}) that evidence the conserved quantities, verifying the results with N-body integrations. Some tests were carried out in the limit of the RTBP.

We analysed the orbital evolution using the different models, and identified the regular and chaotic region in the plane ($\sigma$,$u$) for massive planets. In fact, the phase space structure given by the integrable approximation $\mathcal{H}_{00}$ is not adequate for any condition, and our semianalytical model is more accurate (see Fig. \ref{fig1}). We roughly established a mass limit for the existence of Horseshoe orbits when working with two massive planets depending on the eccentricities and mutual inclinations (see Fig. \ref{fig-limit}).

The analytical $\mathcal{H}_2$ model described correctly the resonant motion up to moderate eccentricities ($e_i\leq 0.3$) and initial mutual inclinations ($J\leq 35^\circ$). It always happens outside the region associated with $QS$ motion. Using the $\mathcal{H}_2$ model, we speeded the orbital evolution by a factor of $\sim 50$ . However, depending on the particular problem, the secular frequencies are overestimated (even $10$ times in our examples). Thus, when working with dissipative forces as tides, Yarkovsky or YORP, the secular effects should be scaled properly in each case. 

The analytical $\mathcal{H}_2$ model was accurate in the case of the general-three body problem with high mutual inclinations, while in the context of the RTBP the semianalytical model $\bar{\mathcal{H}}$ or N-body integrations should be used. 

{We established the location of Lidov-Kozai resonance within the 1:1 MMR. The location of LK resonance centre strongly depends on the mass ratio and on the mutual inclination. The limit for the existence starts from the case of RTBP until $m_2/m_1 \lesssim 0.3$, despite planets with comparable masses force the excitation of the orbits around the equilibrium solution.} 

Thus, when we considered inclined pair of planetary systems, $L_4$, $AL_4$ and $QS$ orbits are the most regulars, and we discover some interesting and very unexpected results for $U$ and $L_3$ orbits. The identified unstable orbits $U$ by \citet{Hadjidemetriou_etal_2009} are, in fact, regular and very stable orbits for pair of Earth-like planets up to mutual inclinations lower than $20\de$. {In the case of inclined systems, contrary to the planar problem, the $L_3$ orbits are very chaotic but bounded. We checked that for $J\lesssim30\de$ the orbits remain stable at least for 50 My.}

The models developed here can be used for a systematic study of the secular dynamics in the coorbital regime with the Solar System planets, and also, with exoplanetary systems. Further work is necessary to study the families of periodic orbits (and stationary solutions). {Moreover, it is necessary further work to characterize the change in the phase space-structure of 1:1 MMR and to give rigorous definitions for the families of periodic orbits in the spatial case and their relationship with planar and also with the restricted case}.

\vspace{-2em}
\bibliographystyle{mnras}
      
\bibliography{library}

\vspace{-2em}
\section*{Acknowledgements}

The authors wish to thank Dr. Beaug{\'e} for his stimulating suggestions and to Dr. A. L. Serra for her valuable comments. We acknowledge financial support
from CONICET/FAPERJ (39593/133325). The numerical simulations have been performed on the local computing resources at the C\'ordoba University (Argentina).

\bsp	
\label{lastpage}

\end{document}